\documentclass{jfm}
\usepackage[utf8]{inputenc}
\usepackage{graphicx}
\usepackage{amsmath,amssymb,amsfonts,amssymb}
\usepackage{subfig}
\usepackage{setspace}
\usepackage{float}
\usepackage{microtype}
\usepackage[normalem]{ulem}
\usepackage{ragged2e}
\usepackage{xcolor}
\usepackage{tabularx}
\usepackage[shortlabels]{enumitem}

\DeclareCaptionJustification{justified}{\justifying}
\usepackage{bm}
\usepackage{hyperref}
\hypersetup{
	hyperindex,
	breaklinks,
	colorlinks=true,
	linkcolor=blue,
	citecolor=magenta,
	bookmarks=true,
	bookmarksopen=true,
	bookmarksopenlevel=2,
	pdfstartpage={1},
	pdfstartview={FitH},
	pdfview={FitH 0},
	pdfauthor={M.-A. Nikolaidis, B. F. Farrell and P. J. Ioannou},
	pdftitle={}}

\usepackage{ifthen}

\def\U{\bm{\mathsf{U}}}

\def\uv{\mathbf{u}}

\def\xv{\mathbf{x}}

\def\C{\bm{\mathsf{C}}}

\def\P{{\bf P}}

\def\A{\bm{\mathsf{A}}}

\def\Fv{\mathbf{F}}
\def\Q{\bm{\mathsf{Q}}}

\def\U{\bm{\mathsf{U}}}

\def\C{\bm{\mathsf{C}}}
\def\Q{\bm{\mathsf{Q}}}

\def\P{{\bf P}}

\def\A{\bm{\mathsf{A}}}

\def\Fv{\mathbf{F}}
\def\Q{\bm{\mathsf{Q}}}

\def\xv{\mathbf{x}}

\def\uv{\mathbf{u}}

\def\st{\sin{\vartheta}}

\def\zhat{\hat{\mathbf{z}}}

\def\xv{\mathbf{x}}

\def\uv{\mathbf{u}}

\def\fv{\boldsymbol{f}}

\def\xv{\boldsymbol{x}}
\def\fv{\boldsymbol{f}}

\def\Fv{\boldsymbol{F}}

\newcommand\redsout{\bgroup\markoverwith{\textcolor{red}{\rule[0.5ex]{2pt}{0.4pt}}}\ULon}

\newcommand{\be}{\begin{equation}}
\newcommand{\ee}{\end{equation}}
\newcommand{\bdm}{\begin{equation*}}
\newcommand{\edm}{\end{equation*}}
\newcommand{\bea}{\begin{eqnarray}}
\newcommand{\eea}{\end{eqnarray}}

\newcommand{\partialf}[2]
{
 \ifthenelse{\equal{#1}{}}{\frac{\partial}{\partial #2}}{\frac{\partial #1}{\partial #2}}
}

\newcommand{\df}{\textrm{d}}

\providecommand\bcdot{\boldsymbol{\cdot}}

\newcounter{saveeqn}%

\def\bt{\tilde{\beta}}

\def\st{\sin{\vartheta}}

\def\xv{\mathbf{x}}

\def\uv{\mathbf{u}}

\def\zhat{\hat{\mathbf{z}}}
\def\nablav{\bm\nabla}

\newcommand{\defn}{\ensuremath{\stackrel{\mathrm{def}}{=}}}
\renewcommand{\equiv}{\defn}

\providecommand\bcdot{\boldsymbol{\cdot}}

\newcommand{\ut}{u_\tau}

\renewcommand{\U}{\mathbf{U}}
\renewcommand{\u}{\mathbf{u}}

\shorttitle{Roll-Streak Formation}
\shortauthor{B. F. Farrell,  P. J. Ioannou, and M.-A. Nikolaidis}

\title{Mechanism of Roll-Streak Structure Formation and Maintenance in Turbulent Shear Flow}

\author{Brian F. Farrell\aff{2}, Petros J. Ioannou\aff{1,2}\corresp{\email{pjioannou@phys.uoa.gr}} \\\and Marios-Andreas~Nikolaidis\aff{1}}
\affiliation{\aff{1}Department of Physics, National and Kapodistrian University of Athens, Athens, Greece
\aff{2}Department of Earth and Planetary Sciences, Harvard University, Cambridge, U.S.A.}

\begin{document}

%
%

\maketitle

\date{\today}

\begin{abstract}
In wall-bounded shear flow the primary coherent structure is the streamwise roll and streak (R-S).   Absent of an associated  instability the  R-S 
has been ascribed to non-normality mediated interaction between the mean flow and perturbations.  This interaction may occur 
either directly due to excitation  of a transiently growing perturbation  or indirectly due to destabilization  of the  R-S by turbulent Reynolds stresses.
A fundamental distinction between the direct and the indirect  mechanisms, which is central to understanding  the physics of turbulence, is that 
in the direct mechanism the R-S is itself the growing structure while in the indirect mechanism the R-S emerges as  a self-organized structure. 
In the emergent R-S theory  the fundamental mechanism is organization by the streak of 
Reynolds stresses configured to support its associated roll by the lift-up process.  This requires
that a  streak  organizes turbulent perturbations such as to produce Reynolds stresses configured to reinforce the streak. 
In this paper we provide detailed analysis explaining physically why this positive feedback occurs and is a 
universal property of  turbulence in shear flow.  DNS data from the same turbulent flow as that used in the theoretical 
 study (Poisseullle flow at $R=1650$) is also analyzed verifying that this mechanism operates  in DNS.

 \end{abstract}


\maketitle
\section{Introduction}
While advances in experiment, simulation and theory continue to be made, the physical mechanisms underlying turbulence in 
shear flows remains incompletely understood.  The problem of shear flow turbulence can be divided into two components: 
transition from the laminar to the turbulent state and maintenance of the turbulent state.  The transition problem is  posed 
by the lack of an inflection in the velocity profiles of boundary layer flows.   Inflections are associated by the Rayleigh theorem 
with existence  of robust instabilities that continue in viscous flows from the inflectional instability of the same velocity profile in an 
inviscid flow.  The problem of robust disturbance growth in perturbation stable shear flows was solved when it was recognized that 
the non-normality of the underlying linear dynamics allows perturbation growth in the absence of exponential instability.  The concept 
of transient growth in shear flow has roots in the classical work of Kelvin and Orr \citep{Kelvin-1887b, Orr-1907}. 
Although three dimensional perturbations in the form of a roll-streak structure were observed in boundary layers 
\citep{Townsend,Klebanoff-etal-1962,Kline-etal-1967,Blackwelder-Eckelmann-1979,RobinsonSK-1991} and related to the nonmodal lift-up growth mechanism \citep{Landahl-1980},
comprehensive observational evidence for the mechanism of nonmodal growth in boundary layer 
flows awaited the advent of DNS at Reynolds numbers  ${\cal O}(1000)$, for which turbulence is maintained, 
and particle image velocimetry (PIV) of turbulent  shear flows.   
The methods of non-normal operator analysis and optimal perturbation theory were first applied  in the context of laboratory shear 
flows to two dimensional disturbances \citep{Farrell-1988a}.  It was believed at the time  that secondary instability of finite amplitude 
two dimensional equilibria were the mechanism of transition \citep{Pierrehumbert-1986, Bayly-etal-1988} and it was  shown that these 
unstable two dimensional nonlinear finite amplitude equilibria could be readily excited by even very small optimal initial 
perturbations \citep{Butler-Farrell-1994}.  However, it became increasingly apparent from observation and simulation that the 
finite amplitude structures associated with transition are three dimensional  and analysis of three dimensional optimal perturbation 
growth followed \citep{Butler-Farrell-1992, Butler-Farrell-1993, Farrell-Ioannou-1993a, Farrell-Ioannou-1993b,Reddy-Henningson-1993,Trefethen-etal-1993,Schmid-Henningson-2001}. These analyses revealed that the optimally growing three dimensional structure is associated with cross-stream/spanwise rolls and 
associated streamwise streaks and is related to the linear lift up mechanism.  The remarkable convected coordinate  
solutions for perturbation growth in unbounded shear flow \citep{Kelvin-1887b} allow closed form solution for the scale 
independent structures producing optimal growth in three dimensional shear flow \citep{Farrell-Ioannou-1993a, Farrell-Ioannou-1993b}.  
These closed form optimal solutions in unbounded shear flow confirm the result found numerically in bounded shear flows that for 
sufficiently long optimizing times streamwise rolls produce optimal energy growth while for short optimizing times the optimal 
perturbations  are oblique wave structures that  synergistically exploit both the two dimensional shear and the three dimensional 
lift up mechanisms producing  vortex cores oriented at an  angle of approximately $60$ degrees from the spanwise direction. 
And indeed, the roll-streak and oblique accompanying structure complex that is predicted to produce optimal growth by analysis 
of non-normal perturbation dynamics of shear flows has been convincingly seen in both observations and simulations 
\citep{Klebanoff-etal-1962, Schoppa-Hussain-2000, Adrian-2007, Wu-Moin-2009,Jimenez-2013lin} and shown to be essentially 
related to the  non-normality of shear flow dynamics
\citep{Kim-Lim-2000, Schoppa-Hussain-2002}.

The most direct mechanism exploiting non-normality to form roll-streak (R-S) structures is to introduce
an optimal perturbation into the flow, perhaps  by using a  trip or other device \citep{Butler-Farrell-1992, Reddy-Henningson-1993, Trefethen-etal-1993}.  A related approach is to stochastically force the  flow which can be analyzed
using stochastic turbulence modeling  (STM) \citep{Farrell-Ioannou-1993e, Farrell-Ioannou-1993-unbd, Farrell-Ioannou-1994b, Farrell-Ioannou-1998a, Farrell-Ioannou-1998b, Bamieh-Dahleh-2001, Jovanovic-Bamieh-2005, Hoepffner-Brandt-2008}.  Because of the non-normal nature of perturbation growth in shear flow,  stochastic turbulence models are  closely related to optimal perturbation dynamics.    In conventional stochastic turbulence models the R-S is envisioned to arise from chance occurrence of optimal or near optimal perturbations in the stochastic forcing \citep{Bamieh-Dahleh-2001, Hwang-Cossu-2010,Hwang-Cossu-2010a,McKeon-Sharma-2010}.   These mechanisms exploit the linear non-normal growth process directly.   
Although the R-S structures could arise from an optimal or near optimal
perturbation occurring by chance in the free stream or instigated by a mechanism such
as a trip or a boundary injection, the ubiquity of the R-S structure in shear 
at low levels of free stream turbulence \citep{Adrian-2007, Wu-Moin-2009} suggests a more
systematic origin and a number of ideas have been advanced to explain the continuous
generation and maintenance of this structure in turbulent shear flow \citep{Schoppa-Hussain-2002, Panton-2001}. 
However, the ubiquity of streak formation suggests, as argued by \citet{Schoppa-Hussain-2000},  that some form of 
instability process underlies the formation of streaks and  
that this instability involves an intrinsic association between the  R-S structure and  associated oblique waves. Such a 
three dimensional instability must differ qualitatively from the familiar laminar shear flow instability 
given that it would necessarily violate the Squire theorem and that extensive search had failed to reveal a candidate instability
in the linearized Navier-Stokes equations.
Nevertheless,
its existence was frequently inferred
from  experiment and simulation e.g.  \citet{Andersson-etal-1999} who concluded
that  the evidence ``..corresponds to some fundamental mode triggered in the flat-plate boundary
layer when subjected to high enough levels of free-stream
turbulence..".  Efforts to identify the mechanism underlying instability of the R-S structure  include
exponential instability mechanisms  \citep{Brown-Thomas-1977} and the Craik-Leibovich instability \citep{Phillips-etal-1996}.
Proposed algebraic growth mechanisms involve a streamwise average torque produced by  interaction of discrete oblique waves \citep{Benney-1960,  Jang-etal-1986}.

In a study of a turbulent shear flow  \cite{Hamilton-etal-1995} noted that
 the Reynolds stresses arising from turbulent perturbations were systematically correlated with the R-S 
 so as to maintain the roll in the manner of a feedback process referred to 
 as the self-sustaining process (SSP).  This correlation was subsequently attributed to the structure of the Reynolds stresses produced by 
 inflectionsl instability of  the streak \citep{Waleffe-1997}, modal critical layer fluxes \citep{Hall-Smith-1991,Hall-Sherwin-2010}
 and transiently growing structures \citep{Schoppa-Hussain-2002}.  
 However,  suppression of unstable modes has been demonstrated to have essentially no effect on maintenance of the R-S providing  
 a constructive proof that inflectional instability 
 is not responsible for the SSP \citep{Farrell-Ioannou-2012,Lozano-Duran-etal-2021}.  
 This left collocation of transiently growing structures as being responsible 
 for providing the roll maintaining torques as suggested by \cite{Schoppa-Hussain-2002}.
 The central physical problem posed by this result is explaining how the perturbation 
 Reynolds stresses are maintained by non-normal growth processes to have the 
 correct amplitude and structure to provide the roll collocated torques required by the 
 instability of the pre-transitional boundary layer R-S as well as the time dependent R-S of the SSP in the turbulent state.

%
The cross stream-spanwise roll structure provides a powerful mechanism for forming streamwise streaks in shear flows when
continuously forced by an oblique wave structure.
However, observation of the spatial correlation of perturbation structures in turbulent flow 
does not support the existence of extensive interfering oblique plane waves.  Nevertheless,
if we observe a turbulent shear flow in the
cross stream-spanwise plane at a fixed streamwise location we see that at any instant  there is a substantial torque from 
Reynolds stress divergence forcing cross stream-spanwise rolls.  The problem is that this torque is not systematic and so 
it vanishes in temporal or streamwise average.  However, in the presence of a perturbation 
streak the symmetry in the spanwise direction is broken and the torque from Reynolds stress divergence  can become organized to 
produce the positive feedback between the streak and roll required to destabilize this structure by continuously and coherently 
exploiting the powerful non-normal R-S amplification mechanism.  The existence of this mechanism for destabilizing the 
R-S in turbulence makes it likely that some dynamical perturbation complex exists to exploit it.    
That this is so was demonstrated by deriving from the Navier-Stokes equations
a system of 
statistical state dynamics (SSD) equations  closed at second order,
 eigenanalysis of which reveals the instability  responsible for destabilizing 
 turbulent shear flow to streak formation \citep{Farrell-Ioannou-2012,Farrell-Ioannou-2017-bifur}.  
 This fundamental instability  underlying  shear flow turbulence had evaded detection in part because 
 it has no counterpart in the linearized Navier-Stokes equations. 
 
 This emergent instability can be understood as a streak formation mechanism 
 in which background turbulence, rather than itself comprising the 
perturbation the non-normal growth of which constitutes the streak, instead 
the stochastically occurring perturbations are organized and collocated by the streak to form optimally 
growing structures with oblique wave form 
that  force the 
cross stream-spanwise roll by inducing a Reynolds stress torque linearly proportional to streak amplitude and collocated with the streak
resulting in an emergent exponential instability of the combined roll-streak-turbulence complex.  

It has been demonstrated constructively using second order SSD analysis of the N-S equations that a dynamical 
perturbation complex exploiting the high non-normality of the R-S plus turbulence complex
exists to  destabilize the pre-transitional boundary layer forming a modal R-S  and that this mechanism persists subsequent 
to transition to maintain the temporally fluctuating  R-S complex in the fully turbulent 
state through a parametric instabiity \citep{Farrell-Ioannou-2012,Farrell-Ioannou-2017-bifur}. However,
the specific mean R-S components and their supporting perturbation structures 
as well as the origin and the mechanism by which these structures produce  torques correctly collocated
to explain the ubiquitous occurrence of the mean R-S 
in wall-bounded shear flow turbulence remains to be fully elucidated and its predictions tested 
against DNS data. In this work we 
analyze the mechanism by which turbulent perturbations are organized by streaks to force and maintain
the R-S 
and compare with DNS to verify that the Reynolds stress 
destabilization mechanism occurs and is consistent with the observed formation and maintenance
of the R-S structure in turbulent Poiseuille flow.

\section{Analysis of the forces responsible for the formation and maintenance of streamwise-mean rolls}

The flow velocity is decomposed into  streamwise  mean, denoted by $\U=(U,V,W)$, with $U$  being 
the mean streamwise velocity, in the $x$ streamwise direction; $V$  the mean 
cross stream velocity in the  $y$ cross-stream direction; and $W$  the mean
spanwise velocity in the  spanwise $z$  direction. 
The velocity deviations from the mean, are   $\u=(u,v,w)$, and referred  to
as the perturbations. The non-dimensionalized equations governing the mean and the perturbations flow and pressure fields $(P,p)$  in a channel take the form
\begin{subequations}
\label{eq:DNS}
\begin{align}
\partial_t\U&+ \U \bcdot \nablav \U  + \nabla P - R^{-1} \Delta \U = - \overline{\u \bcdot \nablav \u} ~,
\label{eq:NSm}\\
 \partial_t\u&+   \U \bcdot \nablav \u +
\u \bcdot \nablav \U  + \nablav p-  R^{-1} \Delta  \u
= -(\u \bcdot \nablav \u - \overline{\u \bcdot \nablav \u})~.
 \label{eq:NSp}\\
&\nablav \bcdot \mathbf{U} = 0~,~~~\nablav \bcdot \mathbf{u} = 0~,\label{eq:NSdiv0}
\end{align}\label{eq:NSE0}\end{subequations}   
with  no slip boundary conditions at the channel walls.  The overline denotes the streamwise average and $R$ is the Reynolds number of the flow.
In the equation for the mean \eqref{eq:NSm} $- \overline{\u \bcdot \nablav \u}$ represents the  Reynolds stress induced forcing of the mean by the
perturbations.    Assuming unit density we do not insist on 
the nomenclature distinction between force and acceleration so that  divergence  of the Reynolds stress induces a streamwise mean force per unit mass
in Eq. \eqref{eq:NSm}, which is independent of $x$, resulting in streamwise velocity component acceleration
\begin{equation}
F_x= -\partial_y (  \overline{uv})- \partial_z (  \overline{uw})~,
\label{eq:Fx}
\end{equation}
and cross-stream and spanwise velocity component accelerations:
\begin{equation}
F_y= -\partial_z(  \overline{v w})- \partial_y ( \overline{v^2})~~,~F_z= -\partial_y (  \overline{v w})- \partial_z (  \overline{w^2})~.
\label{eq:Fyz}
\end{equation}

We want to examine the maintenance of the streamwise mean roll circulation with velocity components  $V$, $W$
and streamwise mean vorticity
$\Omega_x=\partial_y W-\partial_z V$. From \eqref{eq:NSm} we obtain that  $\Omega_x$ is governed by:
\begin{align}
\frac{D \Omega_x}{Dt} = G_x 
+   \frac{1}{R} \Delta_2 \Omega_x~,
\label{eq:MPSI}
\end{align}
with $D/Dt =  \partial_t  +V \partial_y + W\partial_z $
the substantial derivative of the streamwise vorticity under advection 
by the streamwise mean flow $(V,W)$. 
On the RHS  of Eq. \eqref{eq:MPSI} appears
the dissipation term, with the notation   $\Delta_2 = \partial_{y}^2+\partial_{z}^2$
for the two-dimensional Laplacian, and the Reynolds stress roll vorticity forcing term arising from the turbulent perturbations
\begin{equation}
G_x =\partial_y F_z - \partial_z F_y = \left ( \partial_{z}^2 -\partial_{y}^2  \right )  \overline{vw}+
\partial_{yz}  \left ( \overline{v^2} - \overline{w^2} \right )~.
\label{eq:Gx}
\end{equation}
In the  absence of $G_x$ the mean  streamwise vorticity decays
as it then obeys the advection-diffusion equation
\begin{align}
\frac{D \Omega_x}{Dt} = 
 \frac{1}{R} \Delta_2 \Omega_x~,
\label{eq:MPSIa}
\end{align}
which implies 
\begin{align}
\frac{d }{dt} \int  \Omega_x^2 dy dz= -  \frac{1}{R}\int  |\nablav \Omega_x|^2 dy dz~~, ~ \frac{d }{dt} \int (V^2+W^2)  dy dz = -   \frac{1}{R} \int   \Omega_x^2 dy dz~,
\label{eq:MPSIa}
\end{align}
indicating decay in time of  both the square vorticity, $\int  \Omega_x^2 dy dz $, and  the energy of the roll circulation  $\int  (V^2+W^2) dy dz$, 
as noted by \cite{Moffatt-90}. This diagnostic clearly implies that the formation and 
maintenance  of  the streamwise mean roll   requires  a systematic source  of streamwise mean vorticity that can only be provided 
by the rotational component of the Reynolds stress vorticity forcing $G_x$.  
In wall-bounded flows with  cross-stream mean flow shear such rolls would lead to  formation   by the lift-up process  of  low and high speed 
spanwise inhomogeneities 
in the  mean streamwise flow, $U$, referred to as streaks. This streak component is defined as $U_s= U - [U]_z$ where
$[U]_z$ is the spanwise average of $U$. 
These considerations require that a Reynolds stress torque exists to maintain a streamwise 
roll circulation and that a  properly collocated streamwise roll necessarily 
forces a collocated streak so that the R-S is a dynamical 
consequence of  a streamwise mean Reynolds stress torque.  
The dynamical hypothesis of the SSP  \citep{Hamilton-etal-1995,Waleffe-1997}
is that the streak instigates the torque that maintains both the roll and the streak itself  through the lift-up process.
If this is so then explaining how the streak gives rise to its self-sustaining roll-inducing torque 
is the central dynamical problem posed by the maintenance of wall-bounded turbulence. 
Given that the torques are not dependent on modal instability \citep{Farrell-Ioannou-2012}
theoretical explanation of this SSP 
requires identifying how transiently growing structures  produce the required torque.

In order to understand the mechanism of roll formation by perturbation Reynolds stress we must first isolate the component of  $\Fv=(F_y,F_z)$
responsible for forcing the roll circulation. It is clear from \eqref{eq:Gx} that only the rotational part of $\Fv$ is involved in 
roll formation, the substantial part of $\Fv$ that is divergent being cancelled by the immediate development of the pressure field
required by continuity.
The force $\fv$ that remains from $\Fv$  to induce the roll circulation is obtained by a Helmholtz-Hodge decomposition
of the  total Reynolds stress force $\Fv$ into a divergent component,  -$\nablav \varphi$, which acts as a pressure force,
and a non-divergent component $\fv=(f_y,f_z)$, so that: 
\begin{equation}
\Fv= \fv - \nablav \varphi~,
\label{eq:Hodge}
\end{equation}
and $f_y=0$ at the boundaries of the channel.
This Helmholtz-Hodge decomposition in the channel domain is unique \citep{Chorin-1997}. The non-divergent force field $\fv$ 
has components
\begin{equation}
\fv = - \Delta_2^{-1} (  \nablav \times G_x \widehat{\xv} )]~,
\label{eq:f}§
\end{equation}
where $\widehat{\xv}$ is the unit vector in the streamwise direction,  and $\Delta_2^{-1}$  the 
inverse of the two-dimensional Laplacian $\Delta_2 = \partial_{yy}+\partial_{zz}$,
rendered unique by the boundary conditions. The force-field  $\fv=(f_y,f_z)$ 
determines the roll circulation, as it is the  force-field that accelerates 
 the $(V,W)$ velocity field by:
\begin{equation}
\partial_t  V =  f_y = - \Delta^{-1}_2 \partial_z G_x ~~,
~~\partial_t  W = f_z =   \Delta^{-1}_2 \partial_y G_x ~~.
\label{eq:VGx}
\end{equation}  
%

\section{Properties of turbulent perturbation Reynolds stress  in  mean flows without streaks}

Consider 
a perturbation field $\u$ in a channel.  The perturbation dynamics   linearized about the streamwise mean flow $U(y) \widehat{\xv}$, with $\widehat{\xv}$ the unit vector in the streamwise direction,
is governed by 
the equations: 
\begin{equation} 
\partial_t\u +   U \partial_x \u +
( \u \bcdot \nabla ) U~\widehat{\xv}  + \nabla p-  R^{-1} \Delta  \u
= 0~,~~\nabla \bcdot \mathbf{u} = 0~,
 \label{eq:NSpp}
 \end{equation}
 with no slip boundary conditions at the channel walls and periodic boundary conditions in $x$ and $z$. 
 
Because of the homogeneity in both the streamwise and spanwise direction we can identify a component of  the flow field satisfying  \eqref{eq:NSpp} 
with  the form 
\begin{equation}
(u_s,v_s)=\sin(k_z z) e^{i k_x x} (\widehat{u},\widehat{v}),~w_s =\cos(k_z z) e^{i k_x x} \widehat{w},~p_s =\sin(k_z z) e^{i k_x x} \widehat{p} ~,
\label{eq:S} 
\end{equation}
in which the Fourier components $\widehat{u}(y,t)$, $\widehat{v}(y,t)$, $\widehat{w}(y,t)$  satisfy Eq. \eqref{eq:NSpp}.
Perturbations of the form  \eqref{eq:S} comprises a superposition of two  oblique plane waves in the $(x,z)$ plane with wavevectors $(k_x,± k_z)$. 
Perturbations of the form \eqref{eq:S}  are referred to as sinuous about the $z = 0$ axis because the $v_s$ and $u_s$ components 
are antisymmetric while the $w_s$ component is symmetric about $z=0$. 
In order to complete the set of perturbations we choose as companion to the sinuous perturbation field \eqref{eq:S},  the varicose  perturbation field
\begin{equation}
(u_v,v_v)=\cos(k_z z) e^{i k_x x} (\widehat{u},\widehat{v}),~w_v =-\sin(k_z z) e^{i k_x x} \widehat{w},~p_v =\cos(k_z z) e^{i k_x x} \widehat{p} ~.
\label{eq:V} 
\end{equation}
In \eqref{eq:V} the functions  $(\widehat{u},\widehat{v}, \widehat{w}, \widehat{p})$ are the same as those in
Eq. \eqref{eq:S}, as any spanwise shift of the sinuous perturbation field
satisfies the perturbation equations in a spanwise uniform mean flow,  $U(y)$.
Perturbations of the form \eqref{eq:V}, referred to as varicose about the $z = 0$ axis, have  $v_v$ and $u_v$ components symmetric 
and $w_v$ 
component  antisymmetric about $z = 0$.

  \begin{figure}
\centering
\includegraphics[width=0.5\columnwidth]{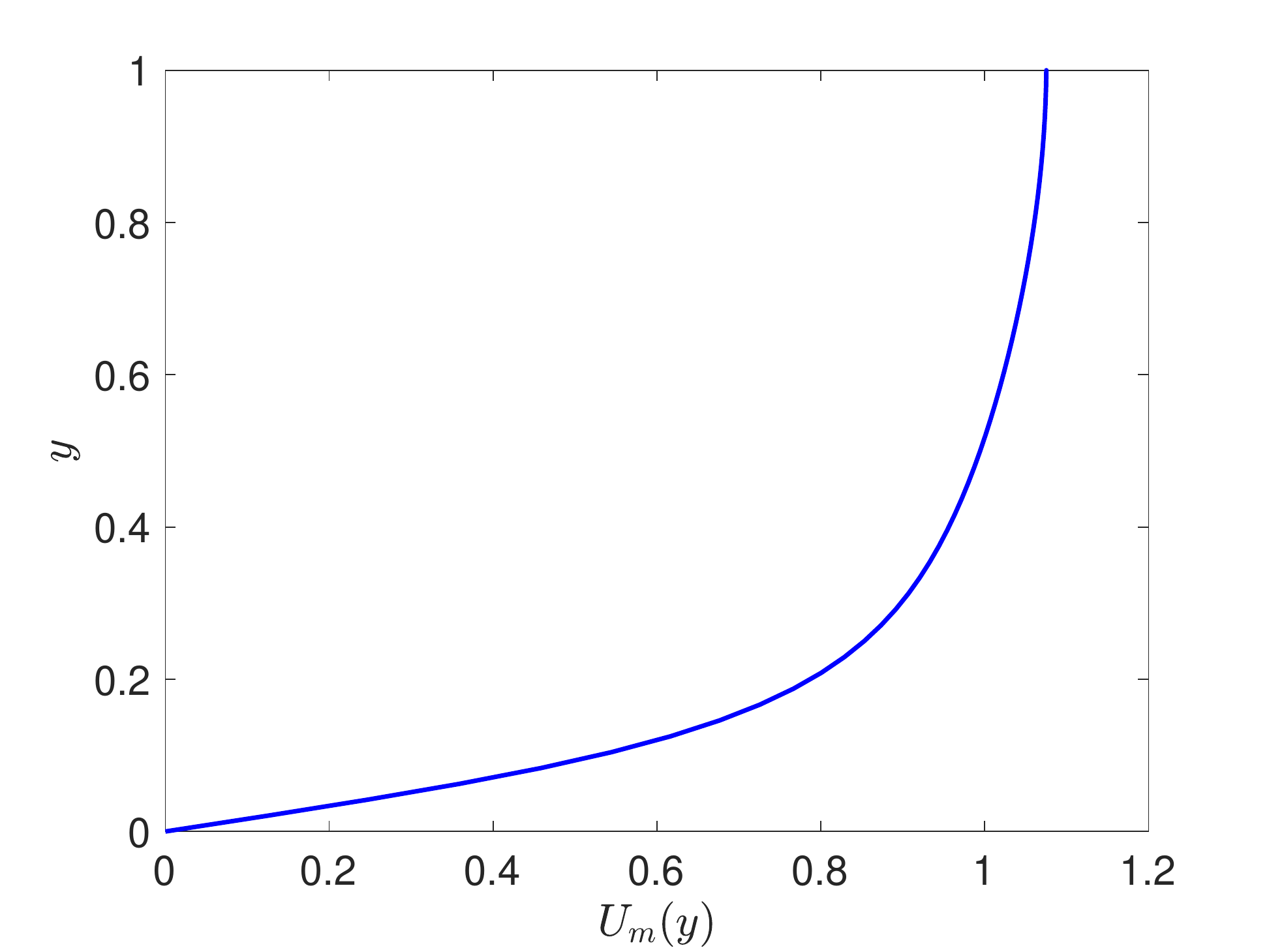}
\caption{The time mean streamwise flow $U_m(y)$ in a turbulent Poiseuille half-channel at $R=1650$.}
\label{fig:Fig1}
\end{figure}

   \begin{figure}
     \centering
\includegraphics[width=0.85\columnwidth]{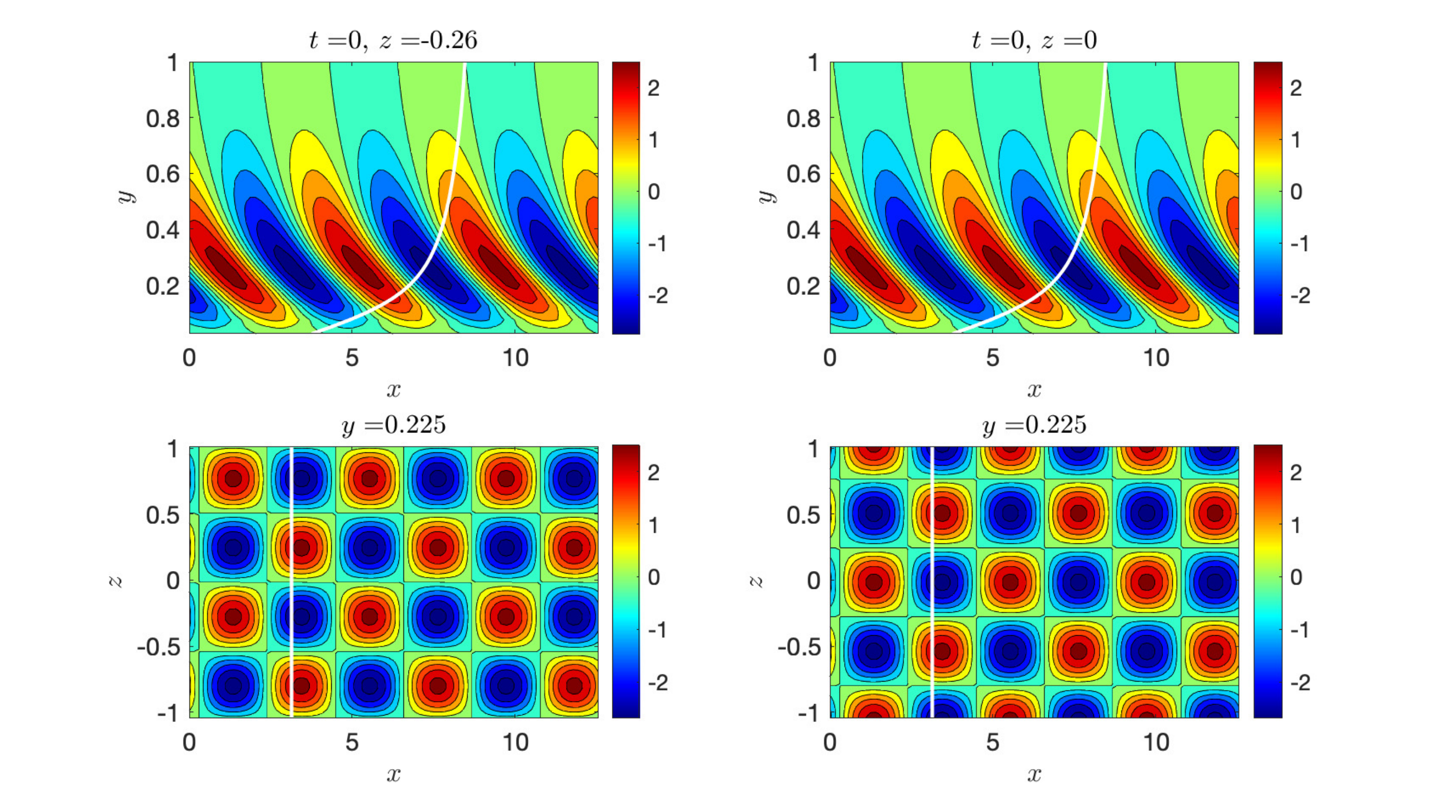}
\caption{Contours of the cross-stream velocity at initial time of the first sinuous (left panel) and first varicose (right panel) $T=10$ optimal with
 $k_x=3 \alpha$.
Both of these optimals achieve energy growth $E_{opt}(T)=23.2$. Top panels: contours of cross-stream velocity at $(x,y)$ cross section 
reveal the expected initial perturbation structure leaning against the 
shear indicative of  growth by the Orr mechanism.  The mean flow in  Fig. \ref{fig:Fig1} is shown by the white line. 
Bottom panels: contours of the cross-stream velocity in the $(x,z)$ plane at the specified $y=0.225$.
The optimals have identical  checker-board sinusoidal structure, with the varicose shifted by a phase of $\pi/2$ in the spanwise direction. }
\label{fig:vf0}
\end{figure}   
   
      \begin{figure}\centering
\includegraphics[width=0.5\columnwidth]{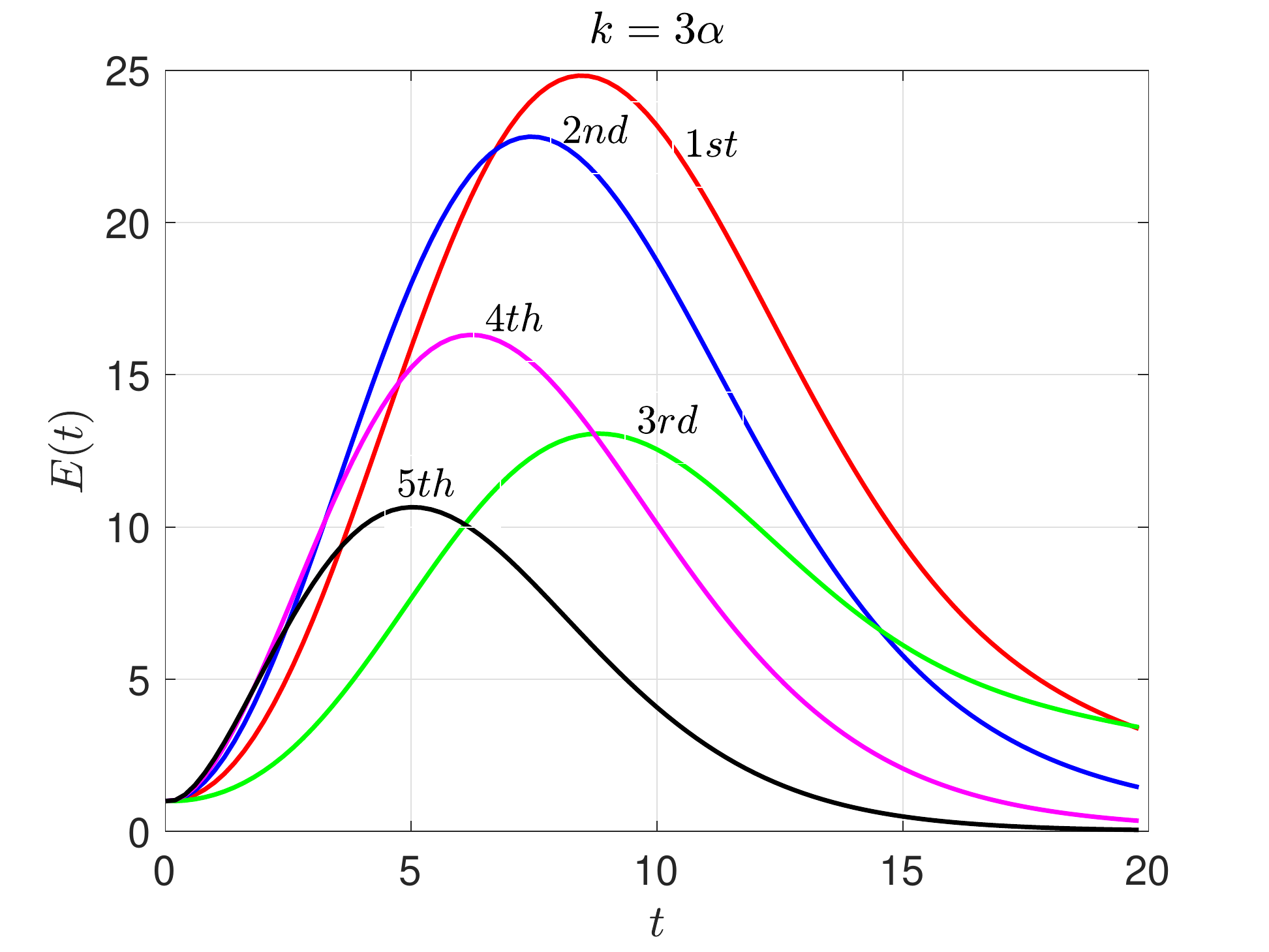}
\caption{The  energy growth as a function of time  of the first five pairs of $T=10$ optimals on the mean flow shown in Fig. \ref{fig:Fig1} for perturbations with $k_x=3 \alpha$ and at $R=1650$. 
The mean flow has no streak and is exponentially stable. }
\label{fig:Eoptf0}
\end{figure}  

Up to this point  the symmetry axis $z=0$ is arbitrary,
but  with the emergence of a streak it can be distinguished  to be  the spanwise location of the streak center. 
When there is  a streak $U_s(y,z)$ in the mean flow, which for theoretical convenience will be made symmetric,  the  sinuous and varicose field components  
will no longer be spanwise translations of each other and 
an asymmetry between these two fields develops.
Quantities derived exclusively from  sinuous perturbations
will from now on be indicated with the subscript $s$, while those of  varicose form will be indicated with the subscript $v$.
\vskip0.05in
We note the following general properties of the Reynolds stresses of sinuous and varicose perturbation fields:\\
\begin{enumerate}[(a)]
\item ~When the mean flow has no streak and is  spanwise homogeneous the sinuous and varicose perturbation fields specified  by \eqref{eq:S} and
\eqref{eq:V} produce 
equal and opposite force-fields $\fv_s=-\fv_v$. 
In this case it is immediate  that the Reynolds stresses of the sinuous
\begin{eqnarray}
&\overline{v_s w_s}= \frac{1}{4}  \sin ( 2 k_z z) ~{\rm Re}(\widehat{v} \widehat{w}^*)~, ~~
\overline{v_s^2 -w_s^2}= \frac{1}{2}  \sin^2 (k_z z) |\widehat{v}|^2 - \frac{1}{2}  \cos^2 (k_z z) |\widehat{w}|^2 ~~,
\label{eq:GXs}
\end{eqnarray}
and the varicose
\begin{eqnarray}
&\overline{v_v w_v}= -\frac{1}{4}  \sin ( 2 k_z z) ~{\rm Re}(\widehat{v} \widehat{w}^*)~, ~~
\overline{v_v^2 -w_v^2}= \frac{1}{2}  \cos^2 (k_z z) |\widehat{v}|^2 - \frac{1}{2}  \sin^2 (k_z z) |\widehat{w}|^2 ~~,
\label{eq:GXv}
\end{eqnarray}
induce equal and opposite torques $G_x = \left ( \partial_{z}^2 -\partial_{y}^2  \right )  \overline{vw}+
\partial_{yz}  \left ( \overline{v^2} - \overline{w^2} \right )$ and therefore equal and opposite force-fields $\fv_s=-\fv_v$.\\

  \begin{figure}
\centering
\includegraphics[width=0.5\columnwidth]{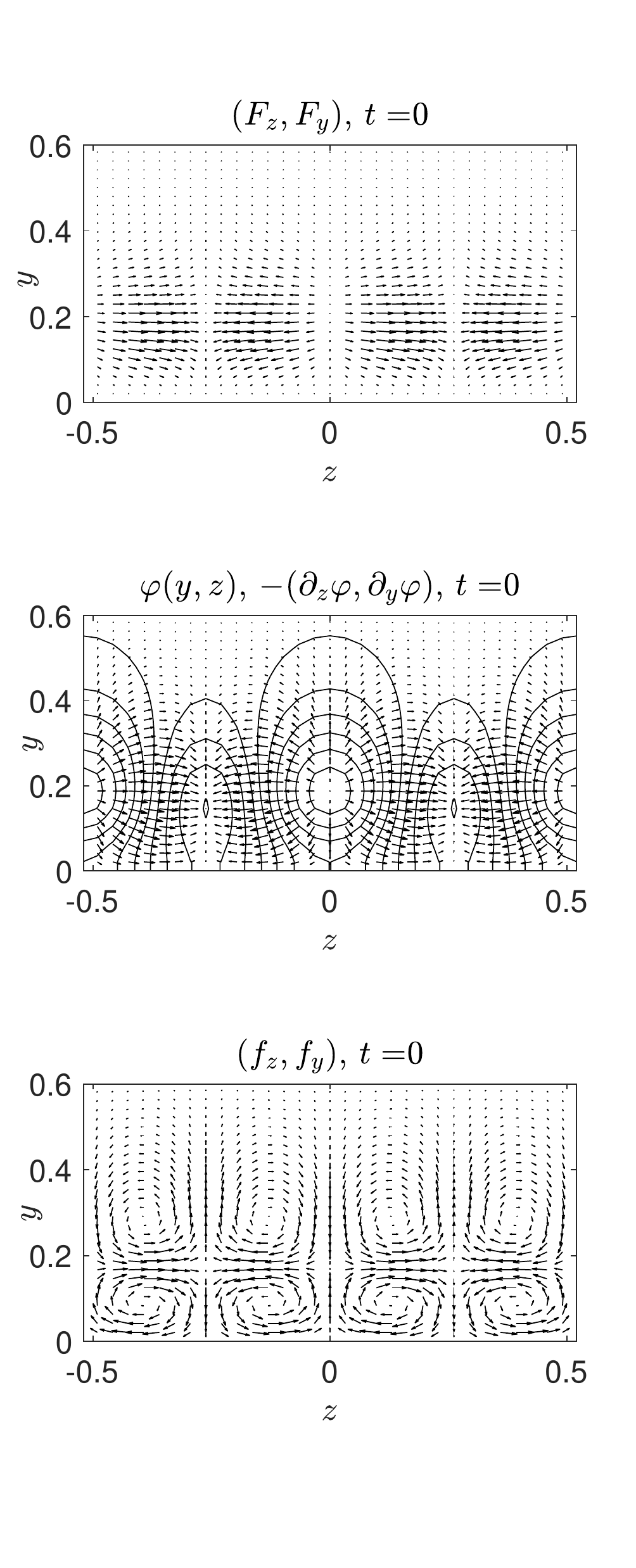}
\caption{Top panel:  vectors  of the total Reynolds stress forcing, $\Fv=(F_z,F_y)$,  given by \eqref{eq:Fyz}, for
the first sinuous optimal in the mean flow shown in Fig. \ref{fig:Fig1}. 
Center panel: contours of the equivalent pressure $\varphi$  of the irrotational part of the Reynolds-stress forcing shown in the top panel.
Vectors indicate the force field associated with this pressure field. Bottom panel: vectors
of the rotational component of the forcing, $\fv=(f_z,f_y)$,  that is responsible for inducing  the streamwise-mean  roll circulation.
Shown is  the  forcing by the $T=10$ sinuous optimal with $k_x=3 \alpha$ at the initial time.  Similar forcing structure characterizes the optimal perturbation  at later times.
}
\label{fig:hodge12S_f0}
\end{figure}

\begin{figure}
\centering
\includegraphics[width=0.5\columnwidth]{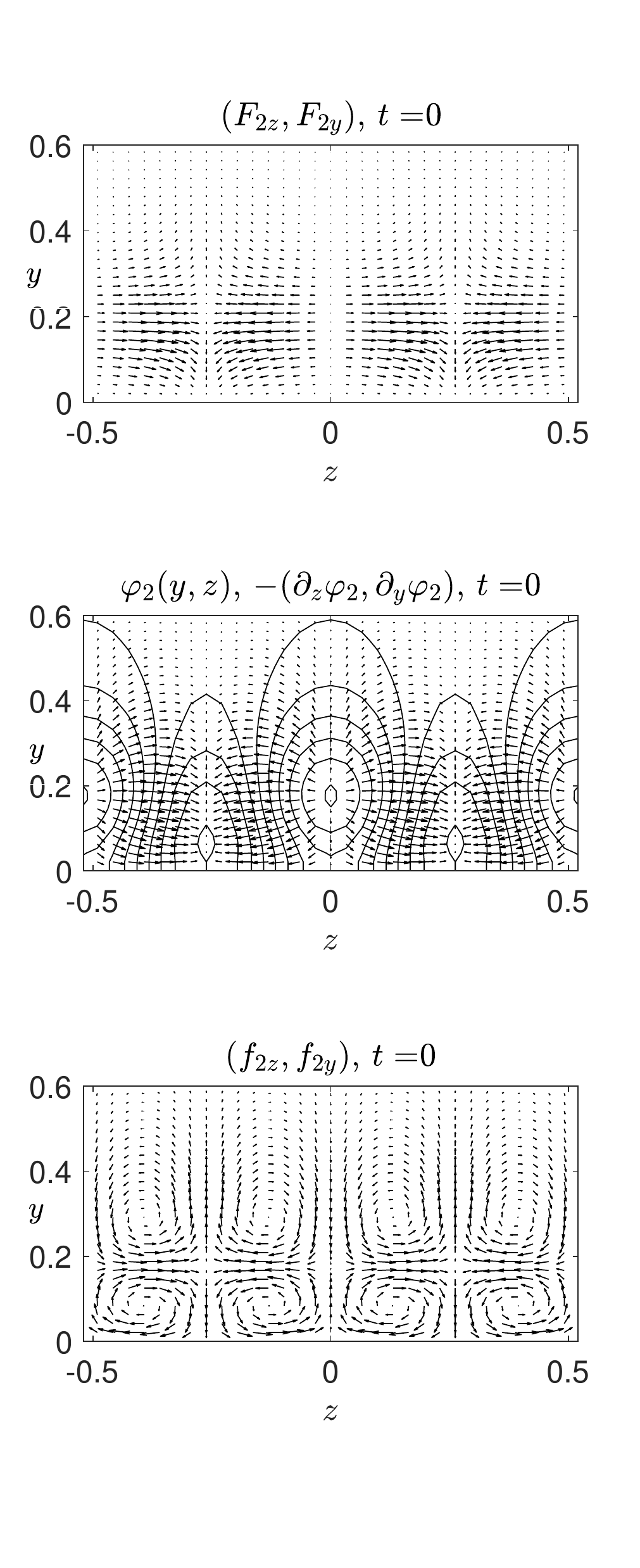}
\caption{Top panel: vectors of the dominant Reynolds stress forcing \eqref{eq:Fyz},
 $\Fv_2=(F_{2z},F_{2y})$, with $F_{2z}=-\partial_z w^2$ and $F_{2y}=-\partial_y v^2$ 
.  Center panel: contours of the equivalent pressure $\varphi_2$  opposing
the irrotational part of the Reynolds-stress forcing of the top panel
with arrows showing the force field associated with this pressure field. Bottom panel: vectors
with  components of the rotational part of the forcing $\fv_2=(f_{2z},f_{2y})$  that is primarily responsible for inducing  the streamwise-mean  roll circulation.
For the  $T=10$ sinuous optimal with $k_x=3 \alpha$ at the initial time.   
}
\label{fig:hodge2S_f0}
\end{figure}


\item~ In mean flows with a symmetric  streak $U_s(y,z)$ exclusively sinuous  or exclusively varicose perturbations 
induce rotational force-fields $\fv$, with $y$ components that are  symmetric about the symmetry axis of the streak, $z=0$,
and $z$ components that are  antisymmetric, i.e. $f_y(y,z)= f_y(y,-z)$ and $f_z(y,z)= -f_z(y,-z)$. This is a general result arising from the symmetry properties of
the velocity fields: under reflection $z \to -z$  exclusively sinuous or varicose fields transform $\overline{vw}$ antisymmetrically 
and $\overline{v^2-w^2}$ symmetrically, implying 
from  Eq. \eqref{eq:Gx} 
that the torque $G_x$ transforms antisymmetrically, $G_x\to -G_x$, and 
from   Eq. \eqref{eq:VGx} that  the force field components  transform as:  $f_y \to f_y$ , $f_z \to - f_z$.\\

\end{enumerate}

Comments:
\begin{enumerate}
{\color{black}
\item ~ The Reynolds stresses  of sinuous and varicose perturbations  \eqref{eq:GXs} and \eqref{eq:GXv} are anisotropic and produce  streamwise-mean
torques  when  $k_z\ne0$,  i.e. when two oblique perturbations with $k_z \ne 0$ and $k_x \ne 0$ interact. 
\item ~Property (a) implies that if a sinuous  perturbation in a mean flow $U_m(y)$ induces a roll circulation, its companion varicose perturbation
induces exactly the opposite roll circulation, and 
consequently a statistically unbiased and uncorrelated  field of perturbations  in a spanwise uniform mean flow $U_m(y)$ can not induce 
a streamwise mean roll circulation. 
However, if spanwise homogeneity is broken  by the purposeful
introduction of a sinuous or varicose
perturbation that breaks the spanwise homogeneity of the perturbation  field  a roll circulation may be forced.
The mechanism of oblique wave interference forcing of roll circulation
was proposed by  \cite{Benney-1960} to be responsible for the
emergence of R-S by interfering
T-S waves in the experiments of \cite{Klebanoff-etal-1962}.
\item ~As discussed, a  perturbation field that is  spanwise homogeneous can not induce  coherent streanwise
roll circulations. However, if the streamwise mean flow is infinitesimally perturbed with a streak perturbation $\delta U_s$,
a homogeneous perturbation field will be distorted  
so as to break the symmetry of sinuous and varicose components and roll circulations will be induced. Remarkably,
at all scales the  perturbation field distorted by a streak results in a rotational forcing configured to amplify the distorting streak perturbation.
Moreover, the roll forcing is not only correlated to  amplify the perturbing streak but also, for 
perturbationally small streak amplitude, to result in roll forcing proportional to the instigating streak perturbation. i.e. 
 $\delta \Omega_x \propto \delta U_s$  leading to exponential instability and the emergence 
 of coherent  finite amplitude streamwise R-S. 
This nonlinear instability is a collective instability that has analytic expression only in the equations for the statistics of the Navier-Stokes equations.
This instability was analyzed in the framework of a second-order closure
of the statistical-state dynamics in \cite{Farrell-Ioannou-2012-doi} and in the full  statistical dynamics of the Navier-Stokes equations in  \cite{Farrell-Ioannou-2017-bifur}.
This is the instability underlying the formation and maintenance of the R-S in turbulent shear flow
that was sought in the linear N-S equations
without success because it does not exist in the state component expression of the dynamics.
}
 \end{enumerate}

\begin{figure}
\centering
\includegraphics[width=1.0\columnwidth]{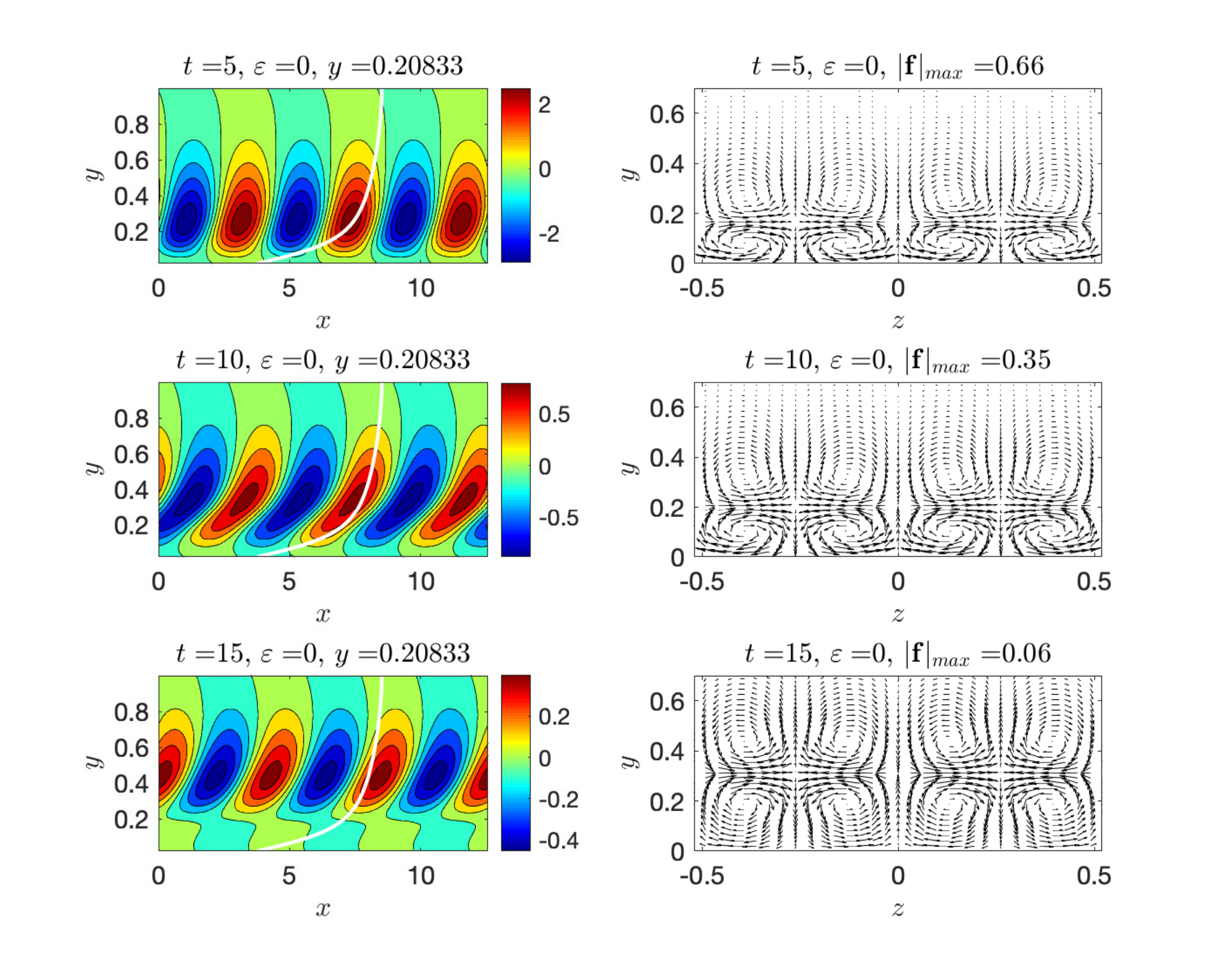}
\caption{Left panels: Contours in the $(x,y)$ plane of the cross-stream velocity of the first sinuous $T=10$ optimal with $k_x=3 \alpha$ 
at times $t=5,10,15$ in the mean flow with no streak. The profile of the mean flow at the cross-section is indicated with the white line.
Right panels: Contours of the low-speed streak and  vectors at the corresponding times of the  Reynolds stress  force 
field that induces the streamwise mean roll circulation in the $(y,z)$ plane. This plot shows that the sinuous optimal at all times induces a roll-circulation
that  tends through lift-up to form  low and high speed streaks. The corresponding varicose perturbation induces the exact opposite circulation. 
The initial energy of the optimal is $E(0)=0.01$.}
\label{fig:opt_sin_f0}
\end{figure}

\begin{figure}
\centering
\includegraphics[width=0.75\columnwidth]{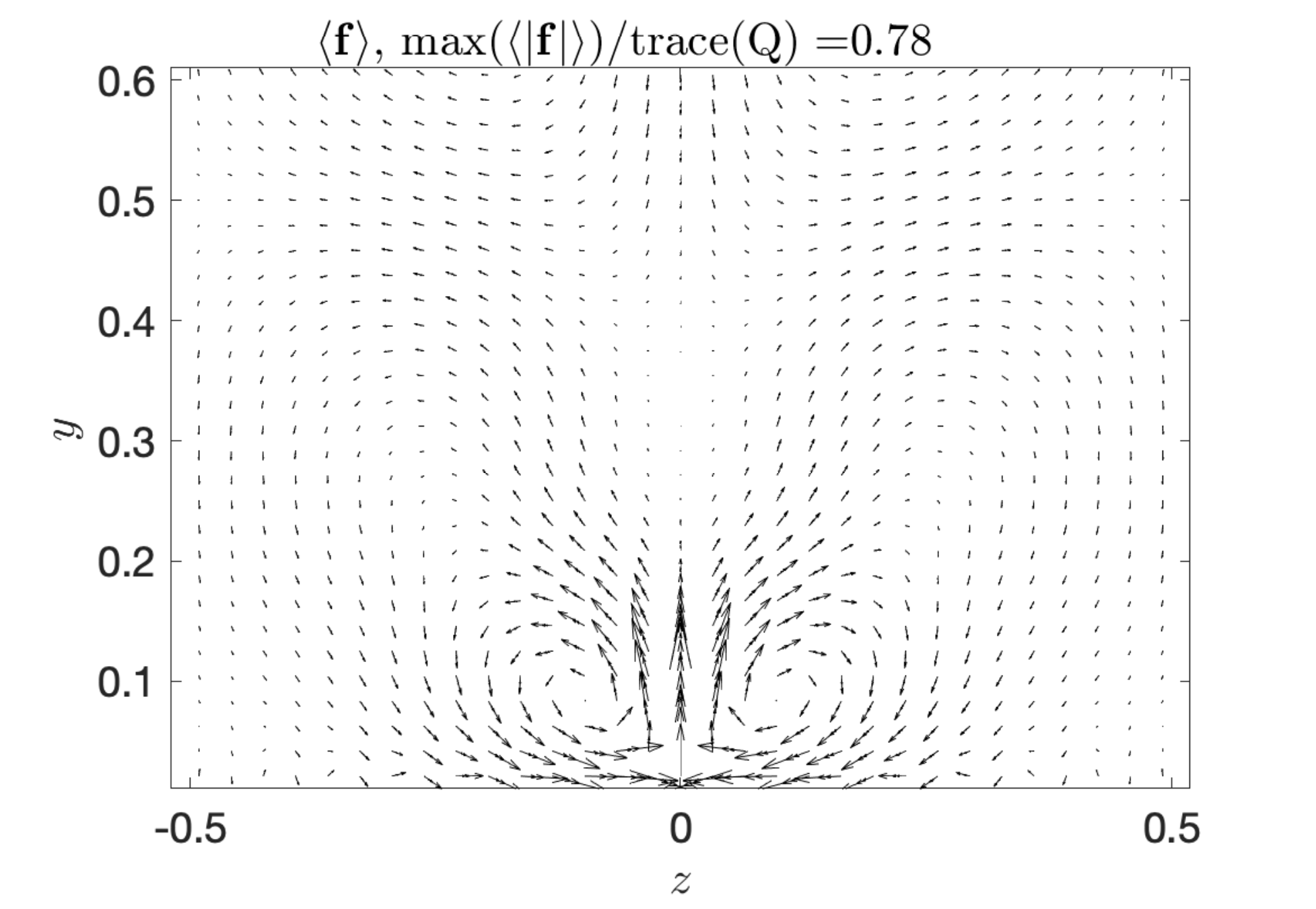}
\caption{Vectors of the rotational component of the ensemble mean Reynolds stress forcing $\langle \fv \rangle =(\langle f_z \rangle ,\langle f_y \rangle)$  of
 the streamwise-mean  roll circulation
for the sinuous components of a stochastically excited perturbation field in the turbulent mean flow shown in Fig. \ref{fig:Fig1}  without a streak.
The varicose part of the Reynolds stress induces in this case the exact opposite circulation so that for the complete perturbation field
the net induced circulation is zero.
The plot is for the $k_x=3 \alpha$ perturbation component. }
\label{fig:stoch_sin_f0}
\end{figure}

\begin{figure}
\centering
\includegraphics[width=0.755\columnwidth]{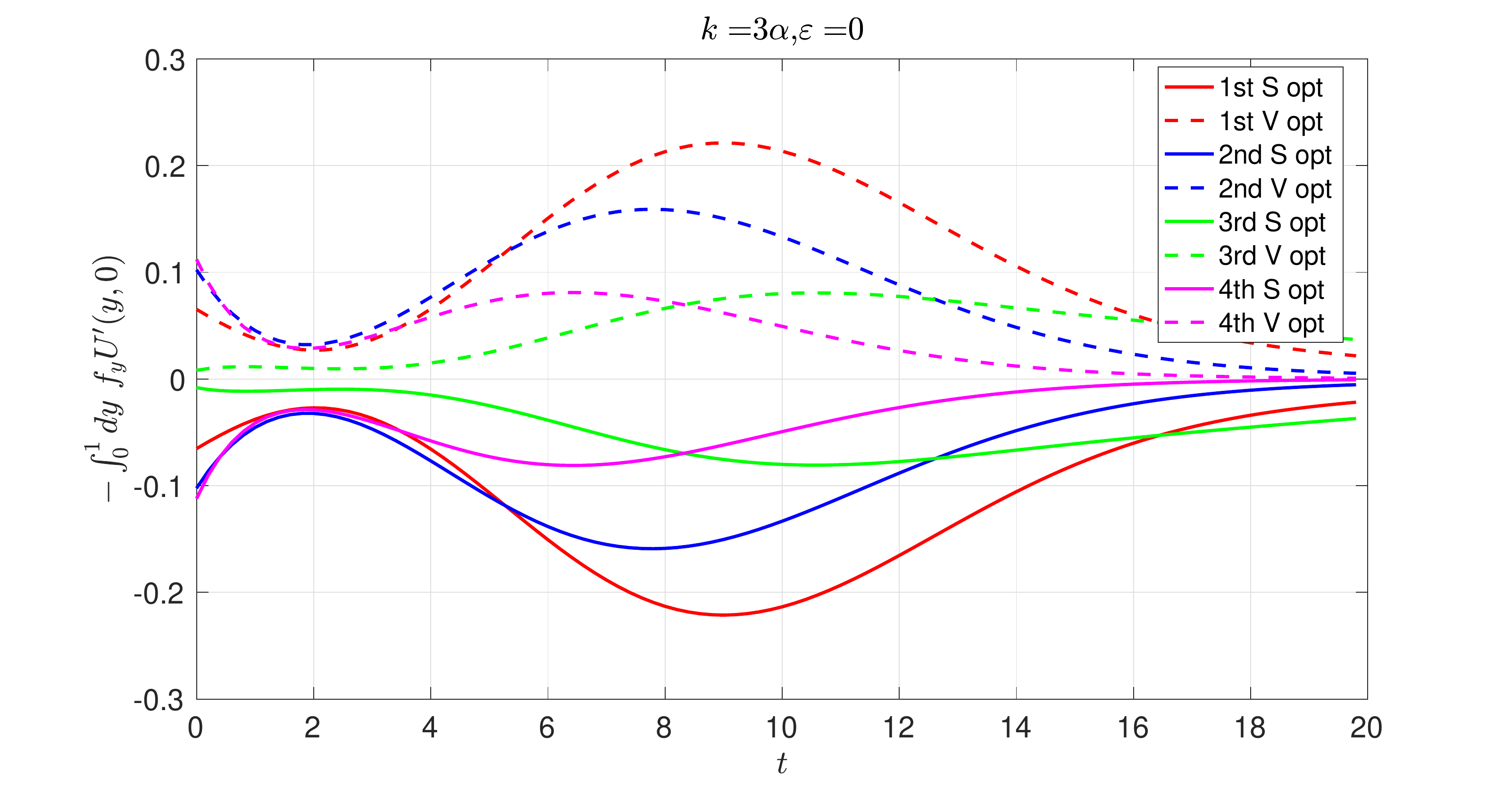}
\caption{Time evolution  of the  acceleration of the streamwise-mean  streak $-\int_0^1 dy~ f_y U'(y,0)$ at $z=0$  by 
first four  $T=10$, $k_x=3 \alpha$ optimal perturbations. The Reynolds stresses of the sinuous optimals (solid) produce  
a positive $f_y>0$, at the symmetry axis near the wall,  and consequently  produce roll circulations that tend to form a low-speed streak,
while the varicose (dashed) induce equal and opposite roll circulations that tend to form a high speed streak.
The induced streamwise acceleration by the optimals is negative at all times for the sinuous perturbations and  positive for the varicose.  
The initial energy of the optimals is $E(0)=0.01$.
}
\label{fig:dU_t}
\end{figure}

   \begin{figure}
\centering
\includegraphics[width=0.5\columnwidth]{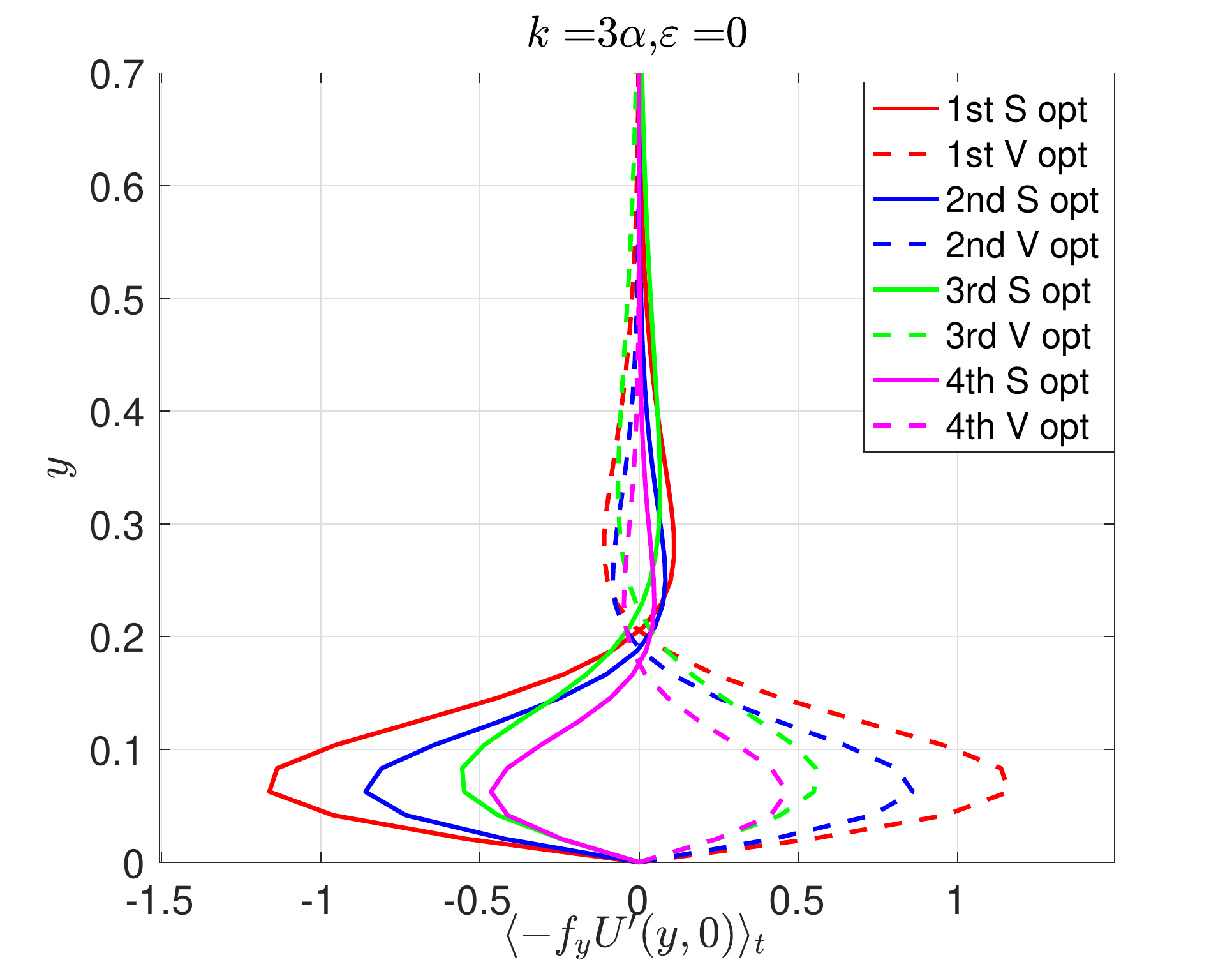}
\caption{The cross-stream distribution  of the time averaged streak acceleration $\langle -f_y U'(y,0)  \rangle_t$ at the streak symmetry axis induced by the 
first four  $T=10$ optimals 
with streamwise wavenumber $k_x=3 \alpha$.  
Shown separately are the contributions from
perturbations, with sinuous (solid) and varicose (dashed) structure.  
The initial energy of the optimals is $E(0)=0.01$.}
\label{fig:dU_y}
\end{figure}

 \begin{figure}
\centering
\includegraphics[width=0.75\columnwidth]{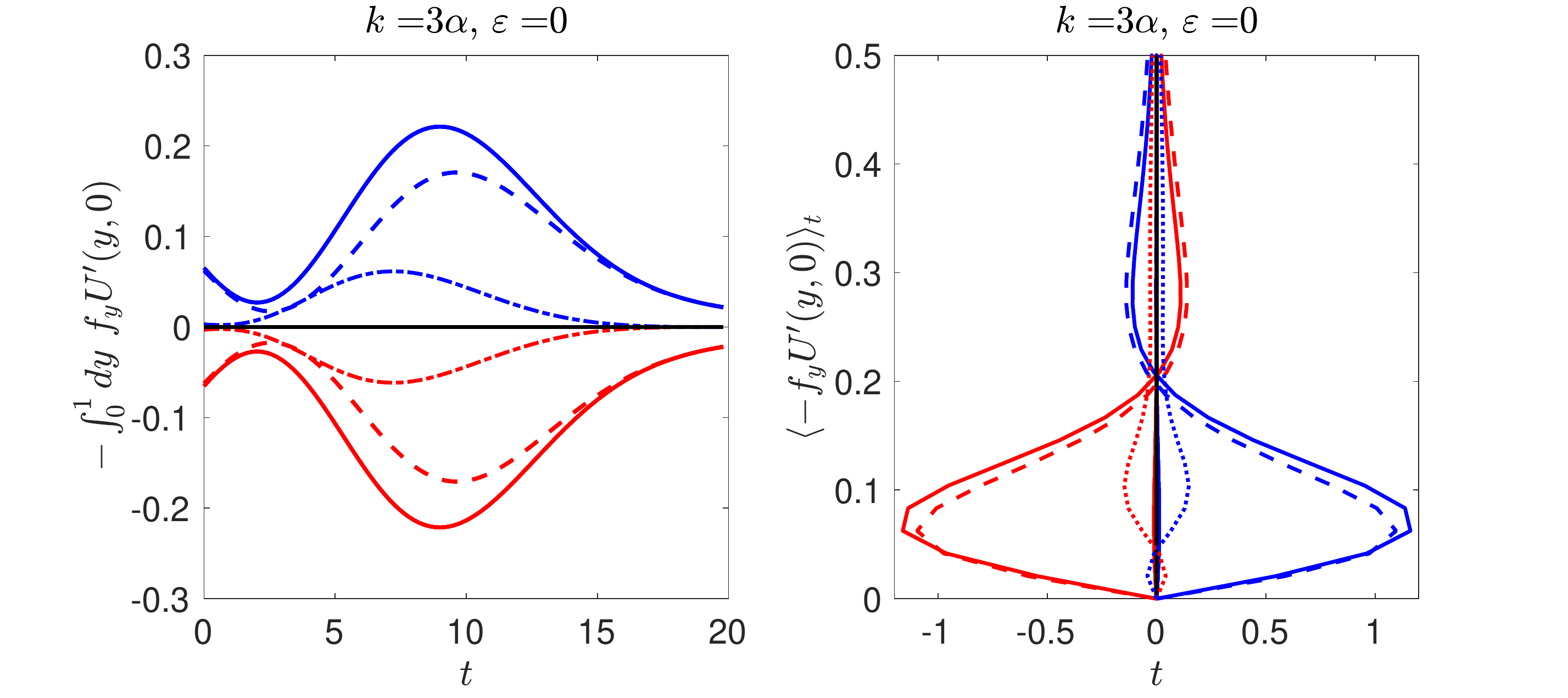}
\caption{Left panel: Time evolution  of the   acceleration of the streamwise-mean streak  $-\int_0^1 dy~ f_y U'(y,0)$ at the symmetry axis $z=0$  induced by the 
first sinuous  (red) and its corresponding varicose (blue) optimal perturbation as shown in Fig. \ref{fig:vf0}. 
Most of the  acceleration is due to the contribution from the $\overline{v^2-w^2}$
Reynolds stress component
(dashed line), the smaller contribution of the $\overline{vw}$
component of the Reynolds stress
is indicated with the dotted line. The net acceleration from this pair of optimals (black line) is zero. Right panel:
The cross-stream distribution  of the time averaged acceleration $\langle -f_y U'(y,0)  \rangle_t$. 
The contribution of $\overline{v^2-w^2}$ (dashed line)
dominates that of
the  $\overline{vw}$ (dotted line).    
The initial energy of the optimals is $E(0)=0.01$ and their streamwise wavenumber is $k_x=3\alpha$.}
\label{fig:dU_sv}
\end{figure}
    
    \begin{figure}
\centering
\includegraphics[width=0.75\columnwidth]{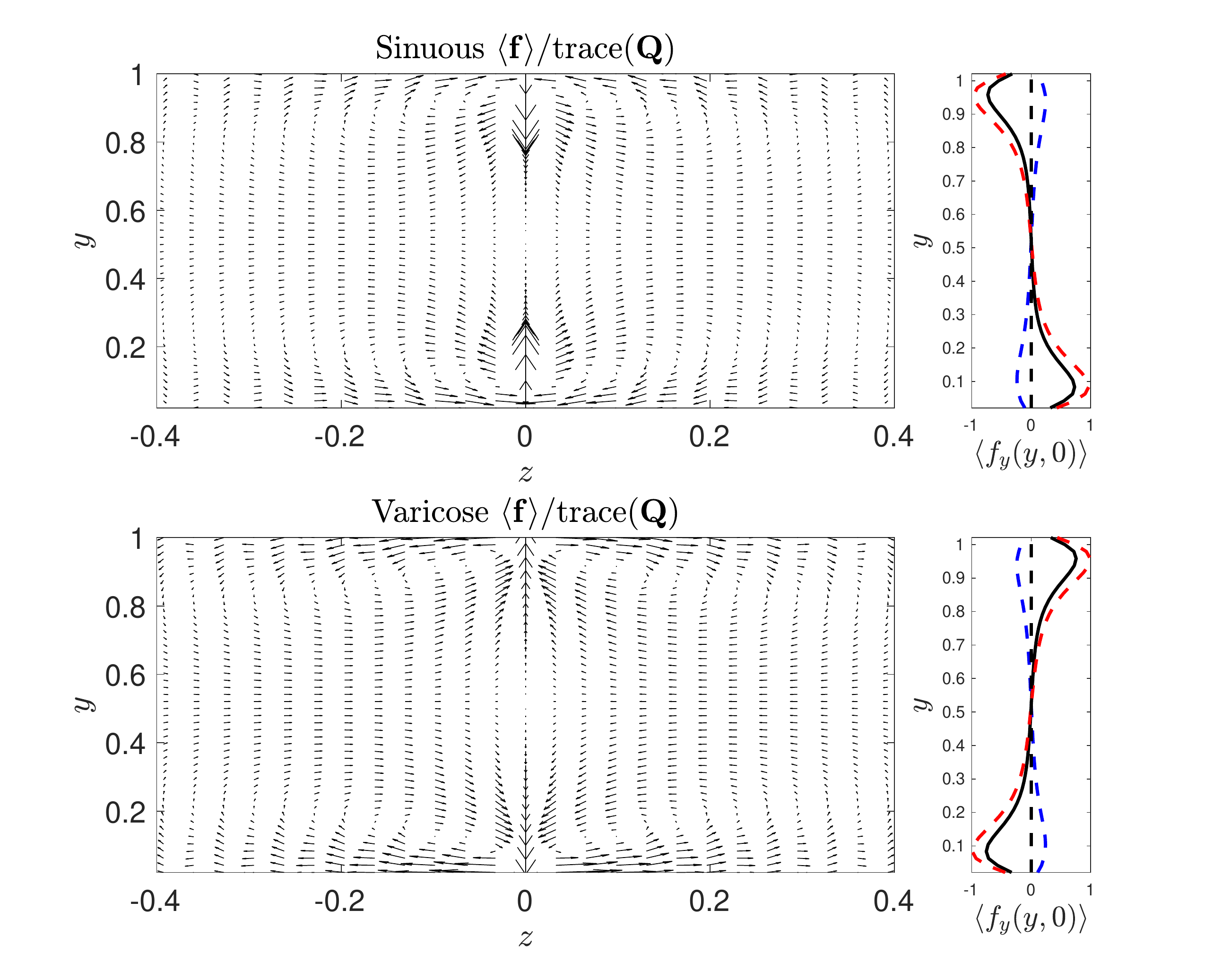}
\caption{Top panels. Left: vectors of the rotational component of the ensemble mean Reynolds stress forcing $\langle \fv \rangle =(\langle f_z \rangle ,\langle f_y \rangle)$  of
 the streamwise-mean  roll circulation
for the sinuous components of a stochastically excited perturbation field with zero mean flow. Right: the cross-stream component $\langle f_y(y,0) \rangle$ 
as function of $y$ at the symmetry axis $z=0$ (black), the dashed-red line is the contribution to $\langle f_y(y,0) \rangle$ from the $\overline{v^2-w^2}$ Reynolds stress, and
the dashed blue line is the contribution from the $\overline{vw}$ Reynolds stress. Bottom panels: the corresponding $\langle \fv \rangle$  for the varicose perturbations, which induces 
the exact opposite circulation so that for the complete perturbation field
the net induced circulation is zero.
The plot is for the $k_x=3 \alpha$ perturbation component, which has been stochastically excited with spatially homogeneous forcing 
in the spanwise. In the cross-stream direction the forcing was weighted by $w=\tanh(y/(5 \delta))+\tanh((L_y-y)/(5 \delta))-1$, where $\delta$ is the cross-stream grid spacing,
in order to reduce the forcing adjacent to the boundaries. The forcing amplitudes have been normalized by the energy input by the stochastic excitation. 
 }
\label{fig:noflow}
\end{figure}

  \section{Roll circulation induced by the Reynolds stresses  of sinuous and varicose perturbations in a spanwise uniform flow}
 
 In the previous section it was shown that the Reynolds stresses of sinuous perturbations  in a spanwise uniform mean flow 
 induce roll circulations that are equal and opposite
to  those induced by their companion varicose perturbations. We
demonstrate in this section that sinuous perturbations with symmetry axis $z=0$
 induce  roll circulations  forming
 low-speed streaks  at the symmetry axis, while  varicose perturbations induce high-speed streaks. 
  
 We choose as  mean flow  the spanwise-averaged turbulent flow $U_m(y)$ 
obtained from a turbulent channel Poiseuille  DNS at $R=1650$ (cf. Fig. \ref{fig:Fig1}).
Our results have been confirmed to be insensitive to this specific choice. This mean flow was chosen in order to facilitate comparison to DNS data.
 The  $(x,y,z)$ flow domain is ${\cal{D} }\equiv[ 0,4\pi] \times [0, 1] \times[-\pi/2,\pi/2]$,
and the perturbations $\u$ evolve according to \eqref{eq:NSpp}, and satisfy periodic boundary conditions in $x$ and $z$ and no slip boundary conditions at $y=0,1$.
The gravest streamwise wavenumber $k_x$  is  $\alpha=2 \pi/L_x$. Our analysis will concentrate on the $k_x=3 \alpha$ component of the flow because in DNS
this component's Reynolds stresses contributed the most in inducing the streamwise-mean roll circulation. The 
same qualitative behavior was obtained for the other streamwise
components of the flow, indicating that the results discussed exhibit insensitivity to scale.

 We consider the Reynolds stresses induced by the optimal perturbations on this mean flow. 
 A $T$-time optimal perturbation  is  the unit energy initial perturbation, $\u(\xv,0)$, that  leads  to the largest energy growth at time $T$, with 
 energy growth:
\begin{equation}
E_{opt}(T) = {\rm max}_{\u(\xv,0)}~ \left (  \frac{\int_{\cal{D}} d^3 \xv ~|\u(\xv,T)|^2}{ \int_{\cal{D}} d^3 \xv ~|\u(\xv,0)|^2} \right )~, 
\end{equation}
(cf. \cite{Farrell-1988a, Butler-Farrell-1992,Farrell-Ioannou-1996a}). 
These perturbations provide an  optimal orthogonal decomposition of the perturbation field according to their growth over time $T$ in this flow. 
The optimal perturbations in this mean flow, which is homogeneous both in the streamwise and spanwise direction, 
are oblique plane waves of the form  $\uv(y,t) e^{i(k_x x + k_z z)}$, characterized by  their  streamwise Fourier wavenumber $k_x$ and  spanwise  wavenumber $k_z$.
These  oblique plane waves can  be combined to form 
symmetric and antisymmetric pairs in the cross-stream  $v$ velocity about $z=0$, which is the symmetry axis of the streak that will be introduced in the sequel.
In this way the optimals of this spanwise uniform flow  can be partitioned into sinuous and varicose perturbations with structure given by Eqs. \eqref{eq:S} and \eqref{eq:V}.  
The structure of the cross-stream velocity of the first pair of sinuous  and varicose $T=10$  optimals  with $k_x=3 \alpha$ are shown in 
Fig. \ref{fig:vf0}. Because of the spanwise translation symmetry
of the mean flow each $(k_x,k_z)$  pair of optimals  with the same cross-stream structure
will have identical energy evolution. For example, the energy growth of the first five $k_x=3 \alpha$
pairs of sinuous and varicose $T=10$ optimal perturbations
are shown in Fig. \ref{fig:Eoptf0}.
 
\begin{figure}
\centering
\includegraphics[width=0.5\columnwidth]{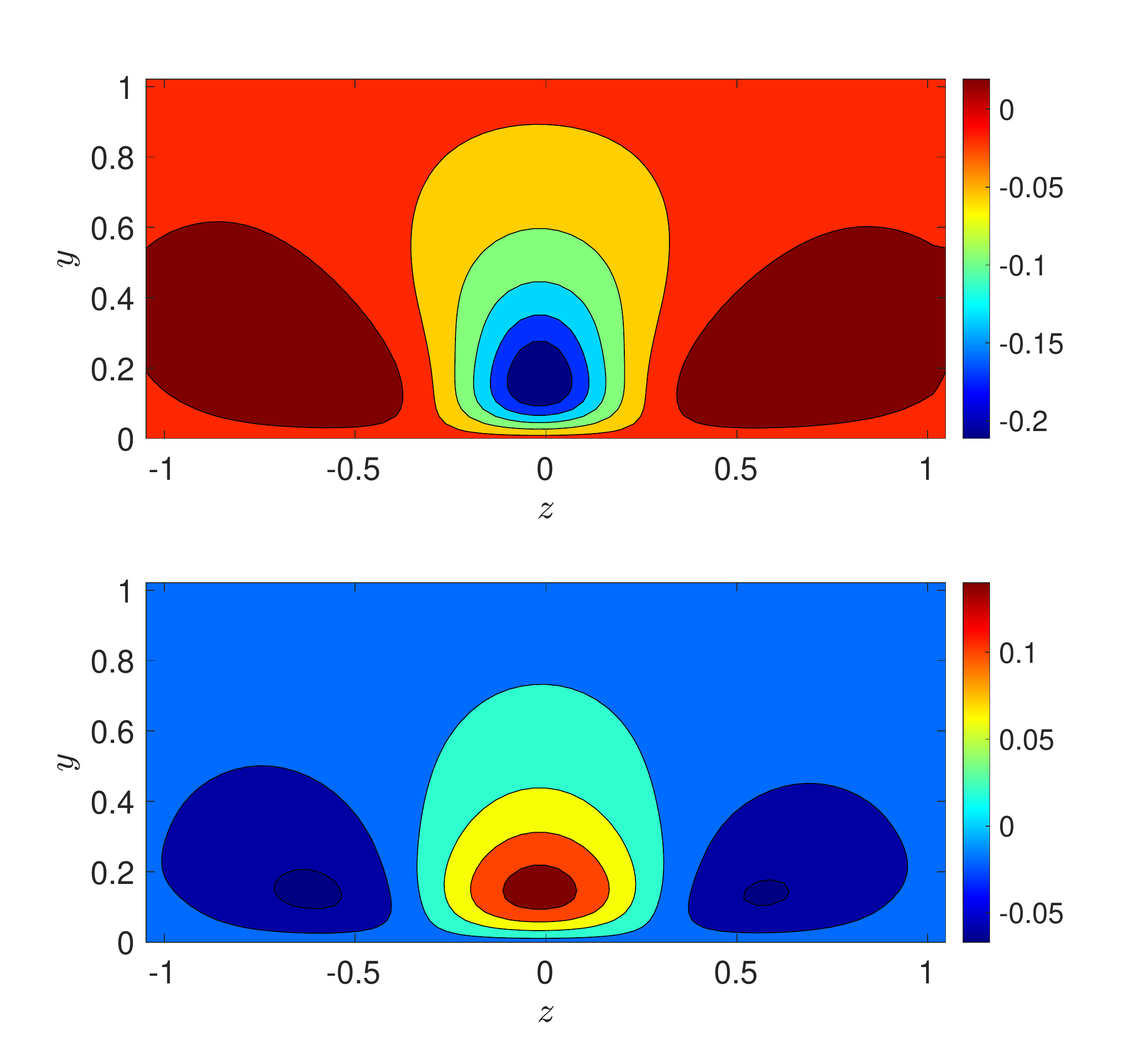}
\caption{ Contours of the velocity of  the equilibrium  low-speed streak (upper panel) and high-speed streak (lower panel) in turbulent Poiseuille flow  at $R=1650$ in a channel with $L_x=4 \pi$ and $L_z=\pi$. The spacing of the contours is $0.04$.}
\label{fig:LH}
\end{figure}

 \begin{figure}
\centering
\includegraphics[width=0.75\columnwidth]{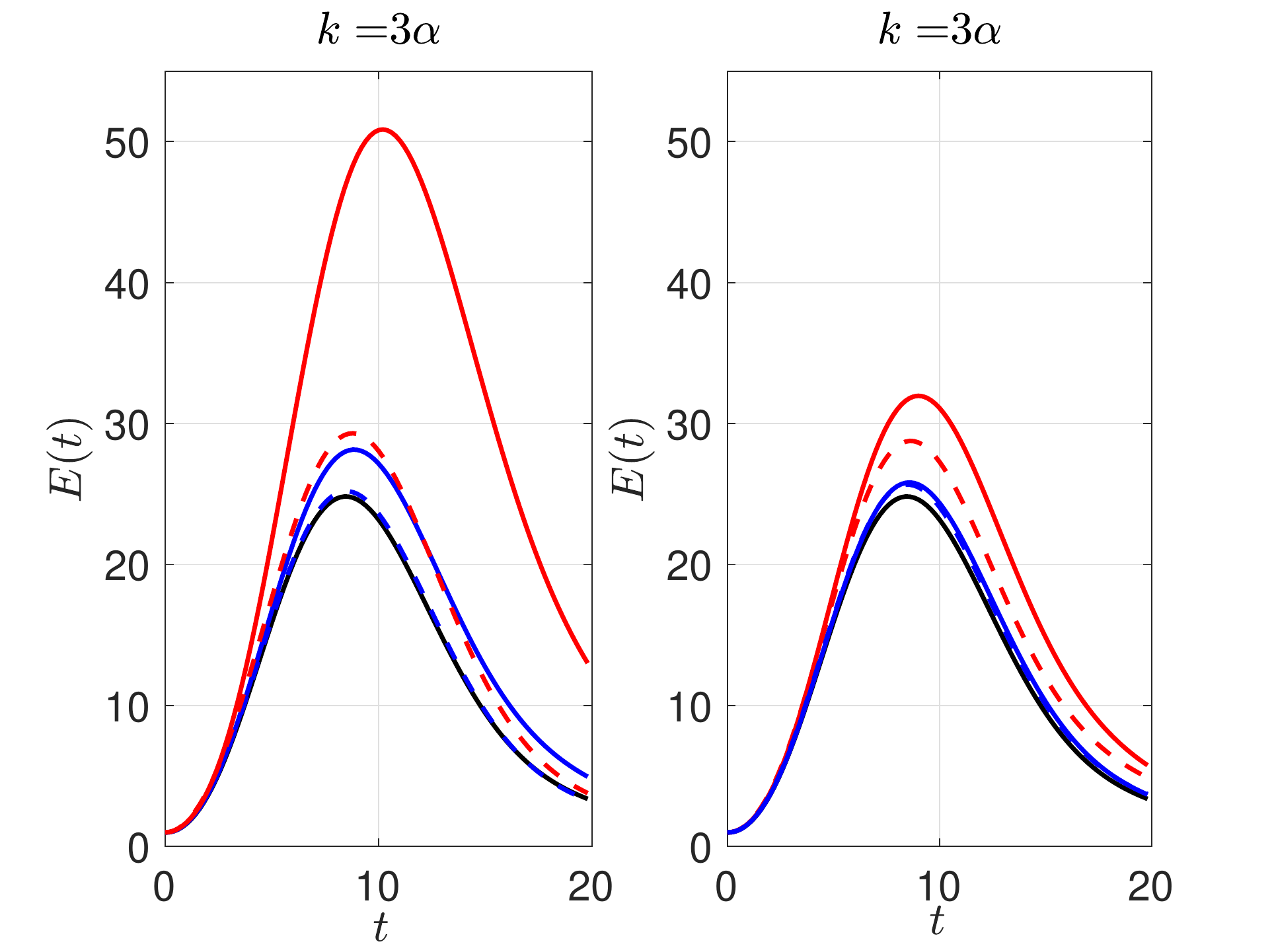}
\caption{Time evolution  of the energy of the first (solid lines) and second (dashed lines) $T=10$ optimals
in a flow with a low speed streak (left panel) and a high-speed streak with the same structure (right panel)  for streak amplitude $\varepsilon=0.4$ and $\varepsilon=1$.
The energy growth of the optimals in the mean flow  with $\varepsilon=\pm 1$ are indicated with red, those  with $\varepsilon=\pm 0.4$ with blue, and those 
with no streak $\varepsilon=0$ with black. 
The first optimal has sinuous structure while the second has varicose structure. This figure shows that the spanwise shear increases the  energy growth
of both perturbations but that the low-speed streak supports substantially greater optimal growth compared to that supported by the  high speed streak.  
Perturbations have  $k_x=3 \alpha$.}
\label{fig:Et}
\end{figure}

  \begin{figure}
\centering
\includegraphics[width=0.75\columnwidth]{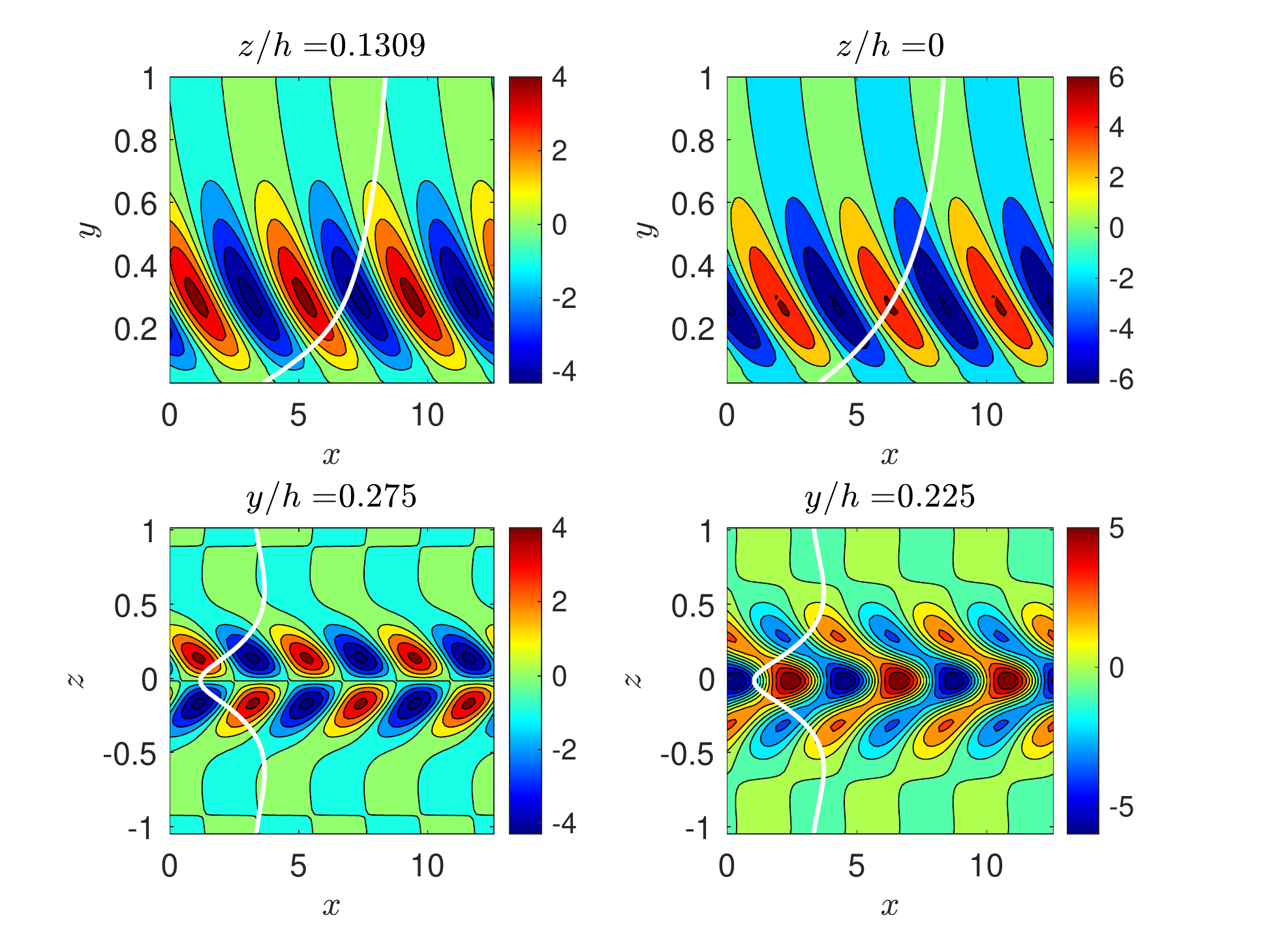}
\caption{ Contours of the cross-stream velocity of the first sinuous (left panel) and first varicose (right panel) $T=10$ optimal for $k_x=3 \alpha$ at the initial time with mean flow
$U(y,z)=U_m(y)+ U_s(y,z)$ in which $U_m(y)$ is the spanwise averaged turbulent mean flow shown in Fig. \ref{fig:Fig1} and $U_s$ is
the equilibrium low-speed streak shown in Fig. \ref{fig:LH} (upper panel).
The presence of the streak centers and confines the optimals at the streak minimum and breaks the symmetry between the sinuous and varicose pairs.
The sinuous achieves the highest energy growth of $50$ at $t=10.4$, while the varicose reaches energy growth $29$ at $t=8.8$.
Top panels: contours of cross-stream velocity in an $(x,y)$ cross section 
show the typical initial configuration  of an optimal perturbation which is that it leans against the 
shear as expected for  growth by the Orr mechanism. The profile of the mean flow is shown by  the solid white line. 
Bottom panels: contours of the cross-stream velocity in the $(x,z)$ plane at  $y=0.275$.
The optimals are obliques waves configured to transfer mean flow energy from the  spanwise shear
to the perturbation. The profile of the mean flow is the solid white line.  The flow is at $R=1650$ in a doubly periodic 
channel with $L_x=4 \pi$, $L_z=2$,
and $\alpha= 2\pi/L_x$.}
\label{fig:optk3L}
\end{figure}

 \begin{figure}
\centering
\includegraphics[width=1.0\columnwidth]{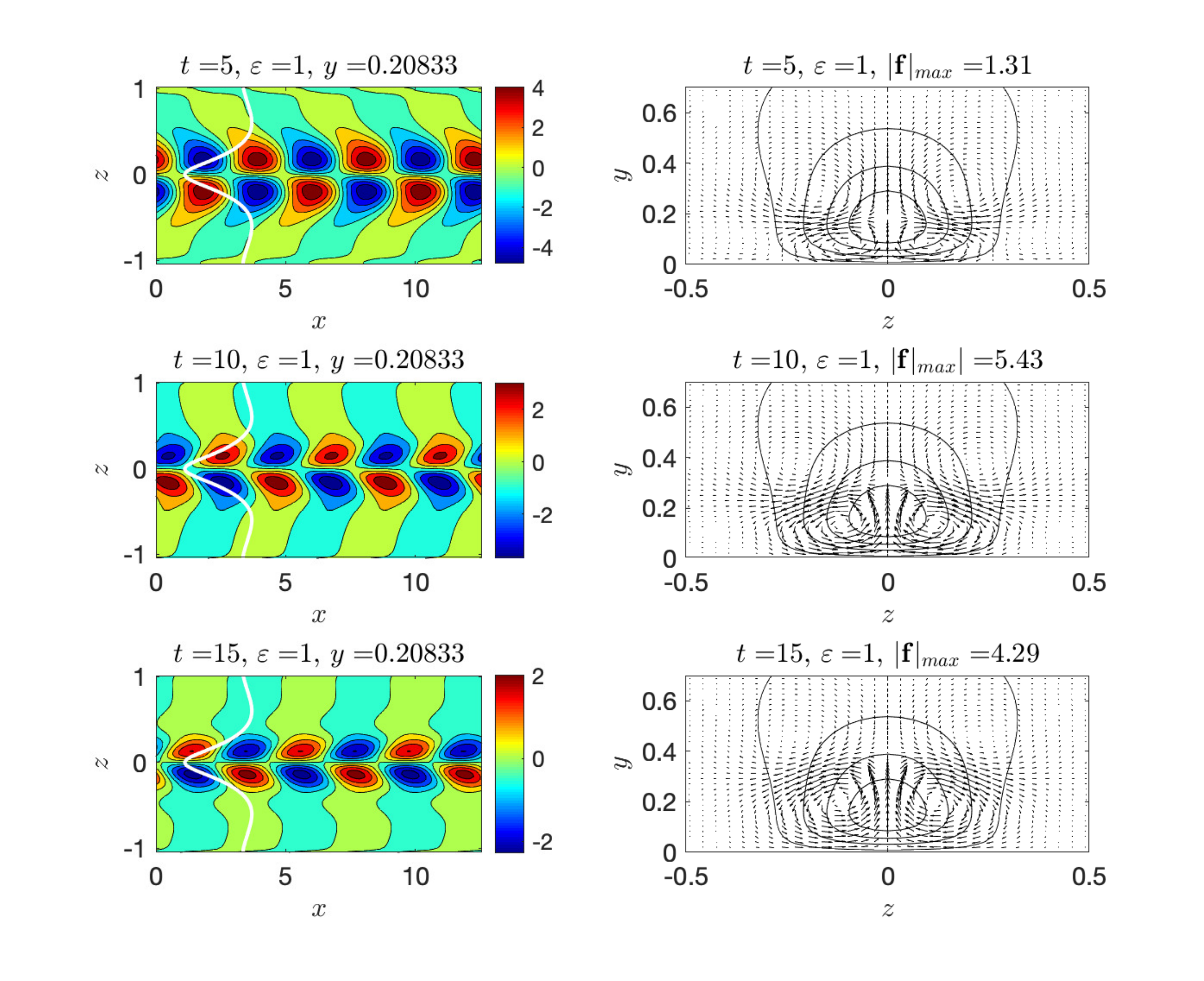}
\caption{Left panels: Contours in the $(x,z)$ plane of the cross-stream velocity of the first sinuous $T=10$ optimal with $k_x=3 \alpha$ 
at times $t=5,10,15$ in the mean flow with the equilibrium low-speed streak. The profile of the streak at the cross-section is indicated with the white line.
Right panels: Contours of the low-speed streak and  vectors at the corresponding times of the  Reynolds stress  force 
field that induces the streamwise mean roll circulation in the $(y,z)$ plane.
The maximum magnitude of the force occurs here at the symmetry axis of the streak. This plot shows that the sinuous optimal at all times induces a roll-circulation
that  tends through lift-up to amplify the low-speed streak. The initial energy of the optimal is $E(0)=0.01$.}
\label{fig:opt_sin_f1}
\end{figure}
 
 \begin{figure}
\centering
\includegraphics[width=1.0\columnwidth]{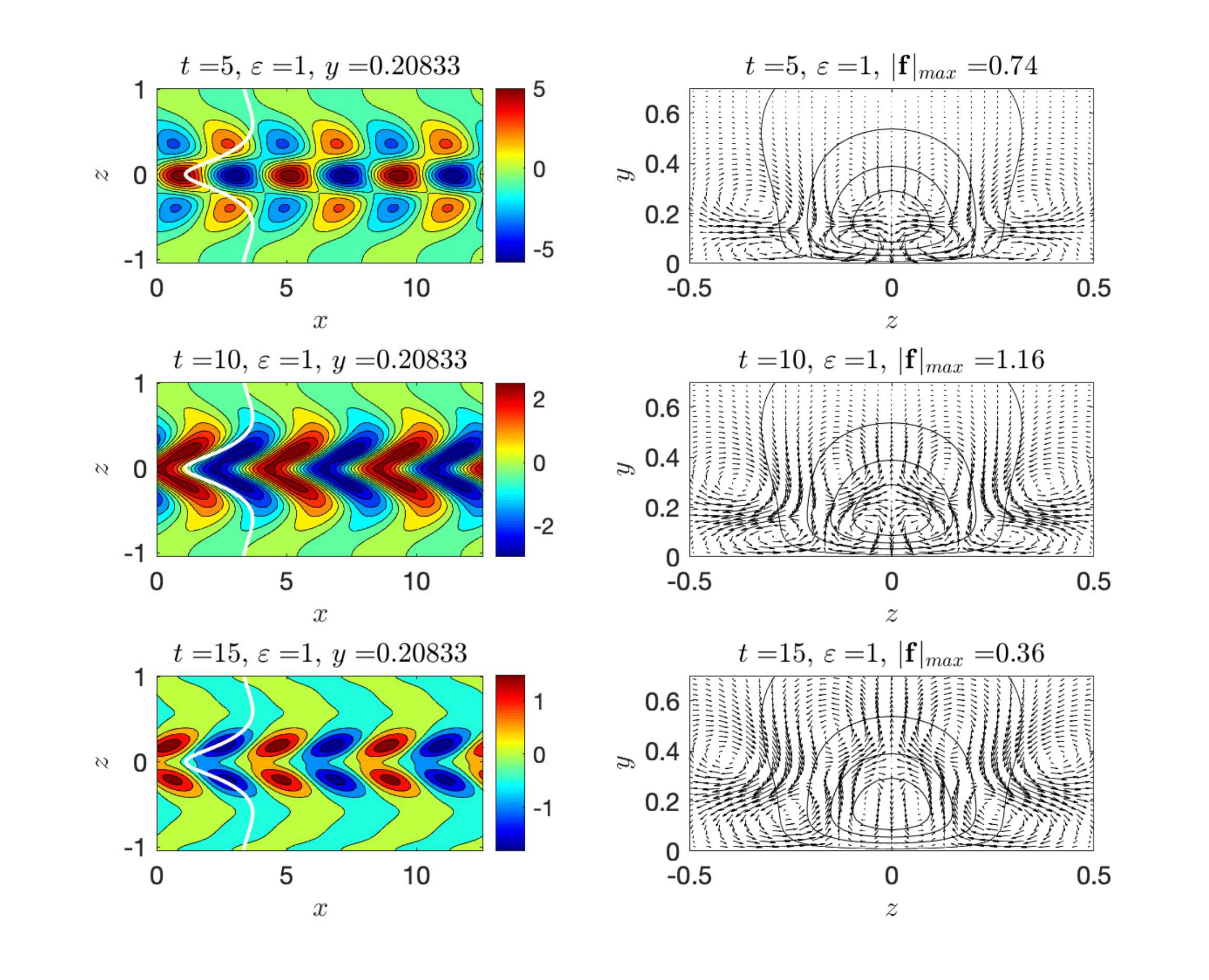}
\caption{As in Fig. \ref{fig:opt_sin_f1}  for the first varicose optimal. This plot shows that the varicose optimal at all times induces a roll-circulation
that  tends through lift-up to destroy the low-speed streak. However the varicose mode induces a weaker circulation than the sinuous  and
the net  circulation from the first sinuous and varicose optimals  results in amplification of the streak as shown in Fig. \ref{fig:opt_tot_f1}. The maximum magnitude 
of the force does not occur  at the symmetry axis of the streak.The initial energy of the optimal is $E(0)=0.01$.}
\label{fig:opt_var_f1}
\end{figure} 
 
 \begin{figure}
\centering
\includegraphics[width=0.75\columnwidth]{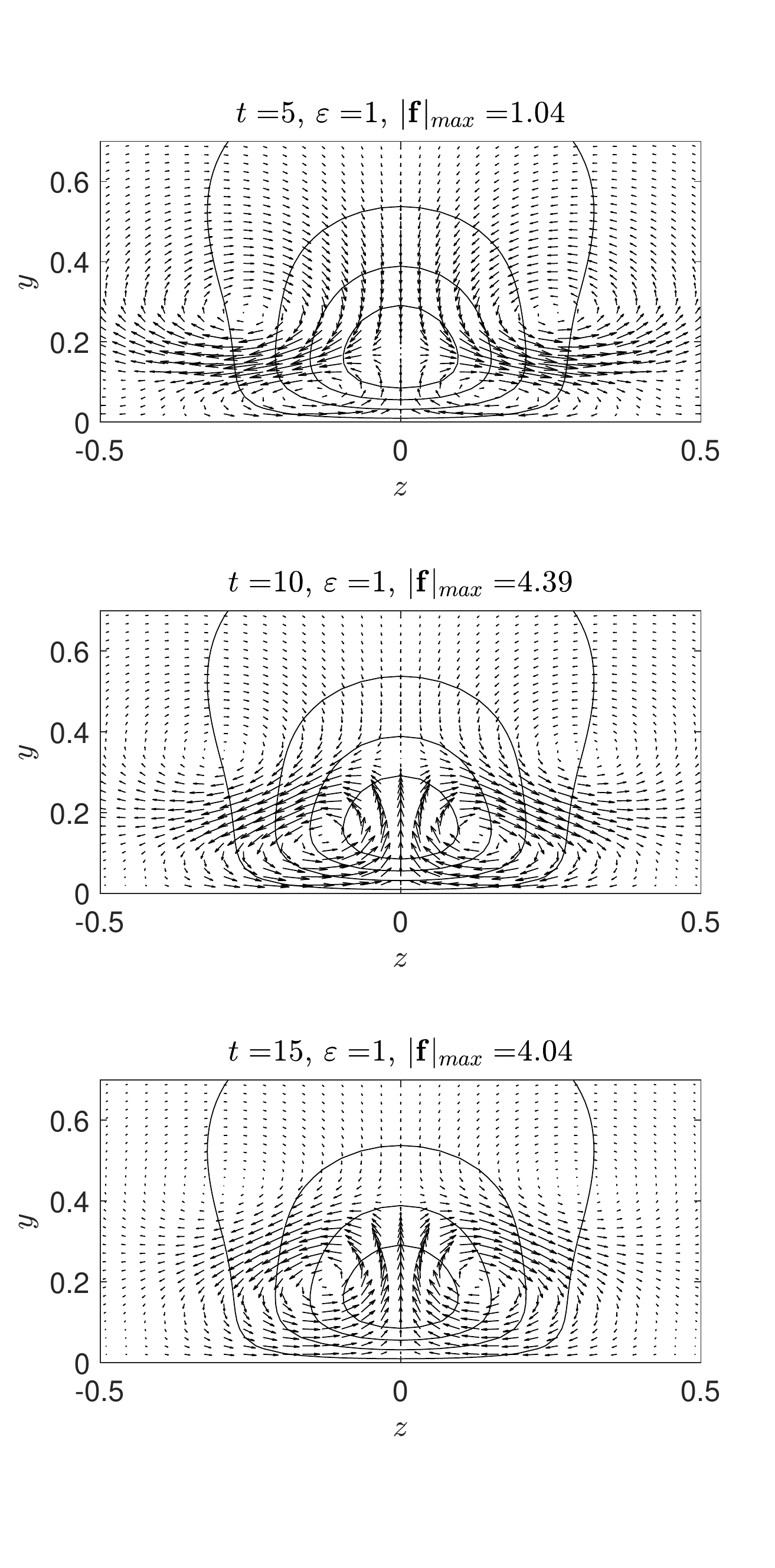}
\caption{As in Fig. \ref{fig:opt_sin_f1}  but vectors of the total force $\mathbf{f}=(f_y,f_z)$ 
obtained by adding the force by the first sinuous optimal    to the force  by the first varicose optimal. The initial energy of the optimals is $E(0)=0.01$.}
\label{fig:opt_tot_f1}
\end{figure}
 
 \begin{figure}
\centering
\includegraphics[width=0.75\columnwidth]{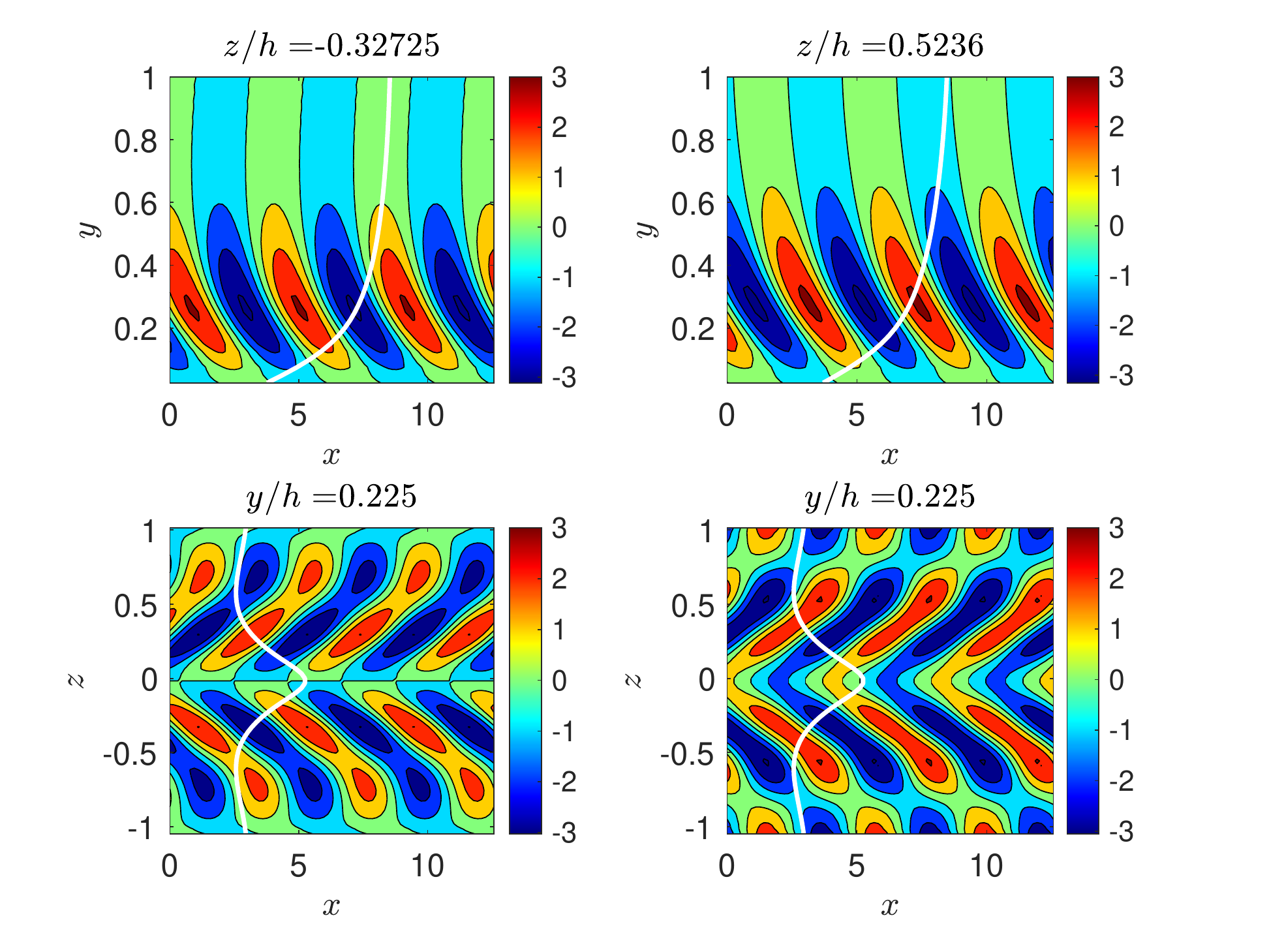}
\caption{As Fig. \ref{fig:optk3L} but for the first sinuous and varicose $T=10$ optimals for the mean flow
$U(y,z)=[U]_z- U_s(y,z)$, with  high-speed  streak  the mirror image of the low-speed streak of Fig. \ref{fig:LH}. 
The first optimal is still  sinuous  achieving energy  growth $32$ at $t=8.4$ while the second optimal is varicose and it  evolves almost identically
with the sinuous  achieving maximum energy growth
$29$ at $t=8.6$. Note that sinuous and the varicose first optimals have almost the same oblique wave structure and that both perturbations are located at the wings of
the streak.  
}
\label{fig:optk3H}
\end{figure}

 \begin{figure}
\centering
\includegraphics[width=0.75\columnwidth]{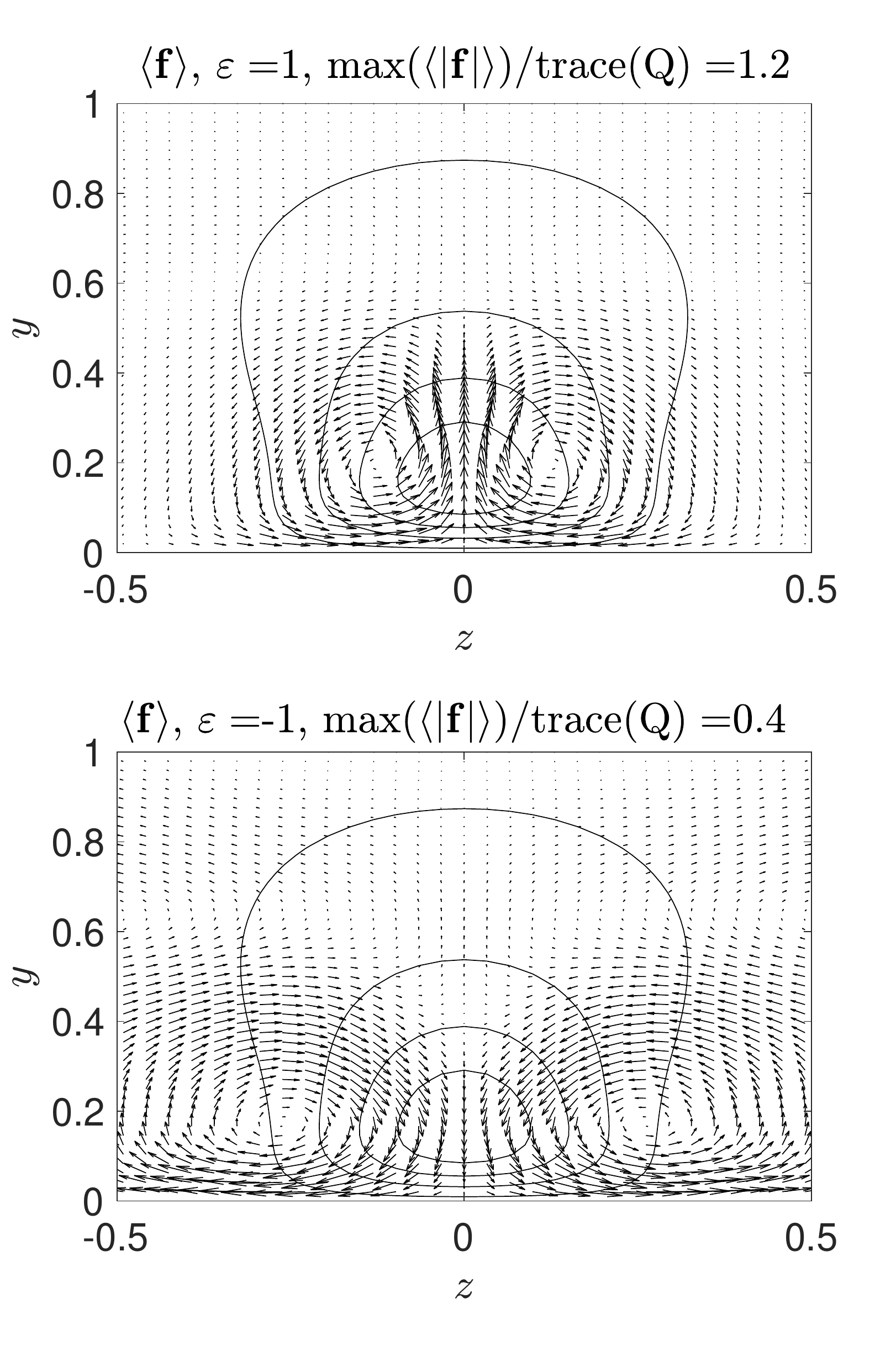}
\caption{Contours of the  equilibrium low-speed streak and vectors of the ensemble mean force field, $\langle \mathbf{f}  \rangle$,
 when the $k_x=3 \alpha$ perturbation field is stochastically excited white in energy in a mean flow  with the  low-speed with $\varepsilon=1$ (upper panel) and 
 with the high-speed streak with $\varepsilon=-1$  (lower panel).
The response is normalized by the energy introduced in 
the flow, ${\rm trace} (\Q)$, where $\Q$ is the forcing covariance in energy coordinates.}
\label{fig:stoch_LH}
\end{figure}


To understand the streamwise-mean roll circulation induced by these optimal perturbations 
we first  plot the divergent total force-field  $\Fv=(F_y,F_z)$ produced by the Reynolds stresses of the sinuous optimal perturbation at the initial time
in  Fig. \ref{fig:hodge12S_f0} (top panel) and perform a Helmholtz-Hodge decomposition of this force-field in
order to determine the characteristics of the residual field $\fv$ that determines the forcing of the roll circulation. 
The total force field $\Fv$ with components
$F_y = -\partial_z \overline{vw}-\partial_y\overline{v^2}$ and
$F_z=  -\partial_y \overline{vw}-\partial_z\overline{w^2}$ is strongly divergent at the symmetry axis $z=0$ 
and is almost completely aligned in the spanwise direction, $\zhat$, i.e.
$\Fv \approx - \partial_z \overline{w^2} ~\zhat$. 
For sinuous perturbations  $-\overline{w^2}$ has a minimum at $z=0$ and therefore the induced $\Fv$ is divergent
in the region about $z=0$, as is evident in Fig. \ref{fig:hodge12S_f0} (top panel). The opposite situation occurs for varicose perturbations.
The divergent component of $\Fv$ is opposed by the  pressure field, $\varphi$, that develops. Contours of the pressure, $\varphi$,
and the resulting force field opposing the divergent component of the Reynolds stress force, $-\nablav \varphi$, are shown in  Fig. \ref{fig:hodge12S_f0} (center panel).
The force from the pressure field is isotropic, and given that  $\Fv$ is divergent along $z=0$,  the pressure force will be directed towards the
channel boundary near the wall and away  far from the wall. This implies that as $\Fv$ has a very small component in the cross-stream direction
the residual $\fv$ will have a strong positive cross-stream component $f_y>0$ at the symmetry axis.
Further from the wall the residual force direction reverses, and the overall force field $\fv$ has quadrupole structure at the symmetry axis,
as seen in Fig.   \ref{fig:hodge12S_f0} (lower panel).

%

Crucial to  the above argument is the dominance of the force field 
induced by the $\overline{v^2-w^2}$ components of the Reynolds stress over the contribution
of the $\overline{vw}$ Reynolds stress. This is shown in   Fig. \ref{fig:hodge2S_f0} 
where the Helmholtz-Hodge decomposition of the force field induced by this Reynolds stress with components
$\Fv_2=(F_{2y},F_{2z}) = (-\partial_y \overline{v^2},-\partial_z\overline{w^2} )$  can be compared to the decomposition of the  full force field 
field  $\Fv$  in Fig. \ref{fig:hodge12S_f0}.
The dominance of the $\overline{v^2-w^2}$ Reynolds stress  is 
maintained during  the  whole evolution of the optimal 
and is the reason, as we will see, underlying the direction of the induced roll-circulation of the sinuous and the varicose perturbations. It should be noted
that  the smaller contribution from the $\overline{vw}$ Reynolds stress changes structure and  sign during the evolution.
We find that dominance of the $\overline{v^2-w^2}$ Reynolds stress 
also characterizes  the  DNS statistics.

 The force field structure  of $\fv$ persists throughout the evolution
of the optimal, as shown in Fig. \ref{fig:opt_sin_f0},  and each developing optimal forces a coherent roll circulation that  through lift-up
forms   a low-speed streak near the wall at the symmetry axis. 
This particular circulation  structure  results from the sinuous form of the perturbations and  is not limited to the first sinuous optimal.
We demonstrate  that by calculating the ensemble mean force $\langle \fv \rangle$ that arises  when all the sinuous components of the flow are equally excited stochastically.
  The ensemble mean statistics are calculated by determining the ensemble mean spatial covariance $\C$ of the velocity components 
which satisfies at statistical equilibrium  the Lyapunov equation
\begin{equation}
\A \C + \C \A^\dagger = - \Q~,
\end{equation}
where $\A$ is the linear operator governing the perturbation dynamics in Eq. \eqref{eq:NSpp},  and
$\Q$ is the  spatial covariance of the delta-correlated stochastic forcing,  chosen here to be the identity in energy coordinates
so that equal energy input is imparted to all degrees of freedom 
(cf. \cite{Farrell-Ioannou-1993e,Farrell-Ioannou-1996a}). The stability of $\A$ ensures that the statistical steady state exists.
From the covariance the ensemble mean Reynolds stresses can be obtained and from them the  ensemble mean 
roll forcing and the lift-up induced streamwise acceleration cf. \cite{Farrell-Ioannou-2012}.

The ensemble mean force $\langle \fv \rangle$ from the sinuous components of the stochastically
excited flow is shown in Fig.  \ref{fig:stoch_sin_f0}. 
This  is the  force field averaged over time  if all sinuous optimals are excited and their Reynolds stress contribution
superposed. Fig. \ref{fig:stoch_sin_f0} shows that the ensemble mean 
Reynolds-stress  force-field from the sinuous perturbations induces roll circulations that tend to form a low-speed streak near the wall at the symmetry axis. Varicose perturbations
induce the exact opposite roll-circulations and is not shown. The resulting roll circulation at the center $z=0$ is similar to the roll circulation induced by the first sinuous optimal.
However the stochastically excited flow results in a roll circulation that is concentrated at the symmetry axis and has  smaller   spanwise wavenumber. This is caused by cancellations
of the induced circulations that occur far from the symmetry axis as the roll circulations from the various sinuous optimals add constructively at the symmetry axis 
and destructively away from it.  
This phenomenon of localization of the roll circulation is  observed during transition to turbulence. Generically, streaks emerge during transition as an exponential instability 
of the interaction of the mean flow and the perturbation field in the background of free-stream turbulence, an instability that has analytical expression in the framework of second order closure
of the statistical dynamics of the channel flow (S3T)  \citep{Farrell-Ioannou-2012-doi,Farrell-Ioannou-2017-bifur}. The streaks that initially emerge are almost sinusoidal
and inherit the spanwise wavenumber of the most unstable mode arising in the S3T closure equations.  But as transition proceeds and the streaks grow the spanwise length-scale increases dramatically
as a broad spectrum  perturbation field  starts being sustained  with its Reynolds stresses contributing to localizing the roll circulation.

To quantify  the development over time
of the rotational force $\fv$ 
generating streaks by lift-up,   
the time development of  quantity $-f_y U'(y,0)$ at the streak symmetry axis (with $U'(y,0)\equiv\partial_y U$ at the symmetry axis)
is shown in Fig. \ref{fig:dU_t} and Fig. \ref{fig:dU_y}. 
This quantity measures the lift-up induced acceleration of the streamwise mean velocity  
at the symmetry axis.
Fig. \ref{fig:dU_t}  shows the time evolution of the net induced streamwise-mean streak $-\int_0^1 dy~f_y U'(y,0)$ at the symmetry
axis  by the first four pairs of sinuous and varicose optimals. The figure shows that at all times the sinuous optimals  induce a low speed streak,
while the corresponding varicose optimals induce  an equal  high-speed streak.  
The time averaged 
cross-stream distribution of the induced streak acceleration $\langle -f_y U'(y,0)\rangle_t$ at the symmetry axis 
by the evolving  first four  pairs of sinuous and varicose optimals
is shown in Fig. \ref{fig:dU_y}.  That the  contribution of the Reynolds stress component 
  $\overline{v^2-w^2}$  dominates that of  $\overline{vw}$ during the development of the first pair of sinuous and varicose optimals is  
shown in Fig. \ref{fig:dU_sv}. 
 
We have seen that key  property of the  sinuous and varicose perturbations to 
 determining the direction of their induced  roll forcing 
 is the dominance of the
 $F_{z}=-\partial_y \overline{vw}-\partial_z \overline{w^2}$ Reynolds stress force component over  
 the $F_{y}=-\partial_z \overline{vw}-\partial_y\overline{v^2}$ component.
 Dominance of $F_z$ is due to dominance of the $\overline{w^2}$ over $\overline{v^2}$, which in turn is due to  the $\overline{v^2}$ fluctuations
being preferentially suppressed compared to  the $\overline{w^2}$ by the solid boundary condition at the wall.  
This asymmetry between $\overline{w^2}$ and $\overline{v^2}$  near a boundary
is a property connected to the presence of the wall and  is not fundamentally dependent on the presence of wall-normal shear or the presence of a
particular type of perturbation, such as an optimally growing perturbation,
other than its being sinuous or varicose.   We demonstrate this by showing that stochastically excited sinuous perturbations in a channel with zero 
mean flow  
produce  a roll-inducing Reynolds stress force field. This resulting roll-inducing 
force field, shown in Fig. \ref{fig:noflow}, has  the characteristic quadrupole structure and is concentrated 
near the walls as is the case for optimal perturbations of sinuous and varicose form. 
When the mean flow has  wall-normal shear the roll-inducing  force field  is concentrated in the shear region,
as shown in Fig. \ref{fig:stoch_sin_f0}, and in this case it is the energy bearing optimal perturbations, 
and their $\overline{v^2}$ field  decaying faster than their $\overline{w^2}$ field
near the wall,
that dominate the perturbation field and are responsible for the structure of the resulting Reynolds stress roll-forcing. 

 \begin{figure}
\centering
\includegraphics[width=0.75\columnwidth]{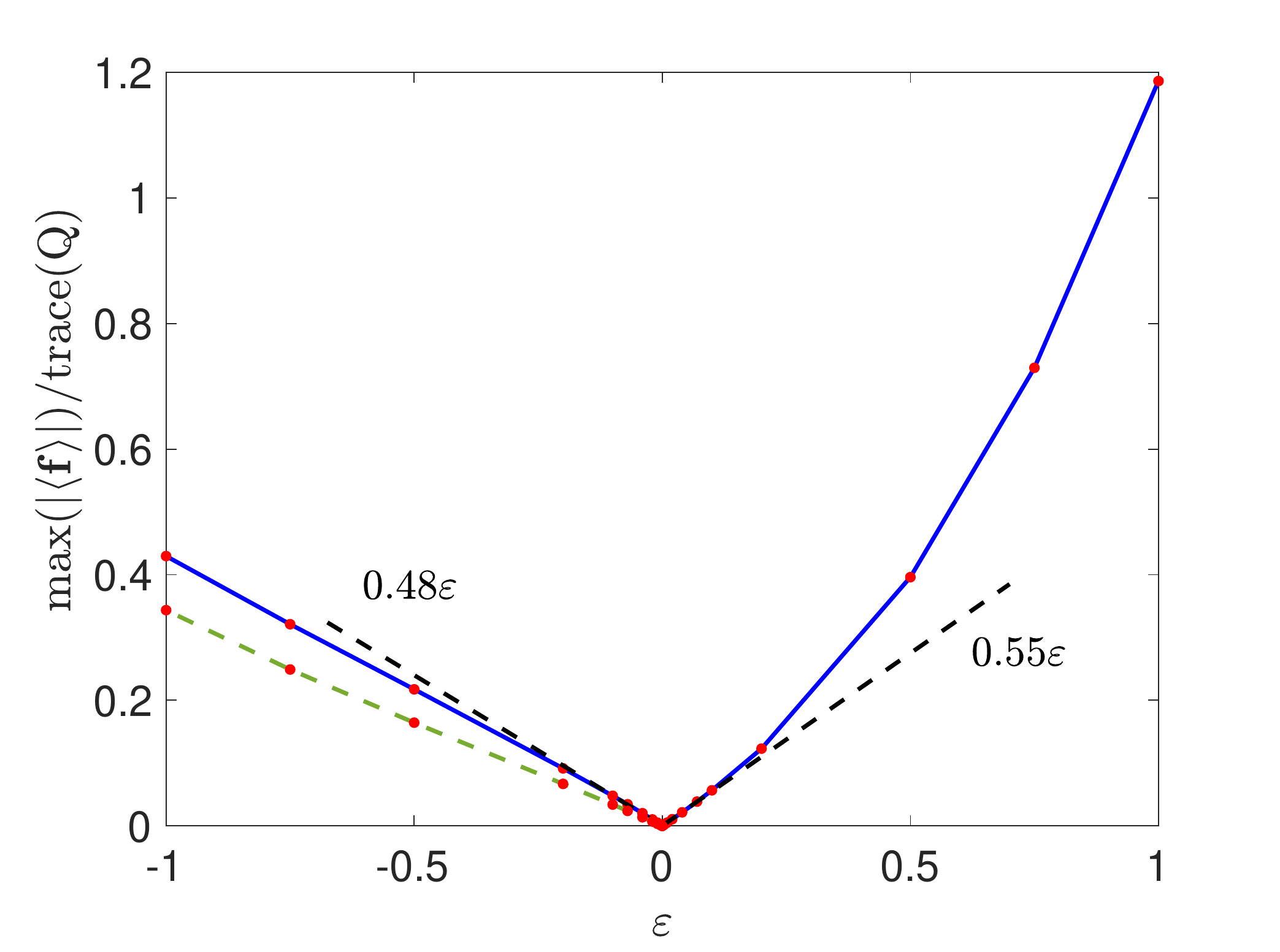}
\caption{The  magnitude of the maximum ensemble mean force $\mathbf{f} = (f_y,f_z)$  produced  by the Reynolds stresses as a function of the amplitude of the streak
$\varepsilon$. This maximum occurs at the symmetry axis of the streak. 
The roll-circulation induced is in a direction to accelerate both the low-speed, $\varepsilon>0$, and the high-speed  streaks, $\varepsilon<0$.\
The dashed green line is the force induced when the high-speed streak has the structure of the equilibrium  high-speed streak shown in Fig. \ref{fig:LH} (lower panel). 
The $k_x=3 \alpha$ perturbations are stochastically excited by a stochastic excitation that injects
${\rm trace}(Q)$ units of energy density per unit time.}
\label{fig:f_epsilon}
\end{figure}

\section{Roll circulation induced by the Reynolds stresses  of sinuous and varicose perturbations in the presence of a  streak}

We have demonstrated that  oblique waves, individual optimal perturbations, 
sums of optimal perturbations and general stochastically forced perturbations 
with sinuous form force roll circulations as do oblique waves, optimals, sums of optimals  and general perturbations with varicose form.  
However, the tendency of sinuous perturbations to form rolls 
configured to force low speed streaks is exactly canceled in a spanwise statistically 
homogeneous flow by the tendency of varicose optimals to form rolls with opposite sign
configured to force high speed streaks.  
It remains to show how the imposition of a perturbation streak breaks 
the cancellation of tendencies in forcing between the  sinuous and varicose components resulting 
in the instability that is responsible for the formation and maintenance of the R-S   in shear flow.

In this section we discuss the Reynolds stresses induced by perturbations in  mean flows with a streak of
the form $U=U_m(y) +\varepsilon U_s(y,z)$,  with $\varepsilon$ a parameter modulating the amplitude
and sign of the streak. The streamwise mean flow $U_m(y)$ is the half-channel time mean  flow 
in  Poiseuille turbulence at $R=1650$, shown in   Fig. \ref{fig:Fig1}, and $U_s(y,z)$ is the equilibrium low-speed streak 
shown in Fig. \ref{fig:LH} (upper-panel) which is  obtained by time averaging the low-speed streaks of the turbulent flow.  With this choice of $U_s$, when $\varepsilon>0$ low-speed streaks are introduced in the flow with the shape of the low-speed streak, and when $\varepsilon<0$ 
 high-speed streaks are introduced that
are   mirror images of the corresponding low-speed streaks. We choose this family of high-speed streaks, instead of those that arise 
from the equilibrium high-speed streak (cf. lower panel of Fig. \ref{fig:LH})  in order to reduce the number of parameters.  Comparisons were made with high-speed streaks 
in the shape of the equilibrated high-speed streak of Fig. \ref{fig:LH}. The results presented here were found to be
robust and not sensitive to this detail.     

Upon introduction of a spanwise symmetric streak the perturbations were found to  localize about the center of the streak.
The sinuous and varicose perturbations continue to induce roll circulations  predominantly from the 
Reynolds stress associated with $\overline{v^2-w^2}$, as previously described, and as previously described these
roll circulations  induce respectively low and high-speed streak perturbations. 
However,
the introduction of the streak creates a crucial difference which is 
an imbalance between the  sinuous and varicose perturbations. With no streak
the net circulation induced by corresponding sinuous and varicose perturbations cancel. 
In the presence of a low-speed streak
the sinuous perturbations are favored over the varicose and the induced  net  roll circulation  tends to reinforce the low-speed streak, while  
in the presence  of a high-speed streak the varicose perturbations are
favored resulting in a net roll circulation that tends to reinforce the high-speed so that in both cases the Reynolds-stress 
induced circulation reinforces  the pre-existing streak. We will demonstrate this dynamics for the first pair of optimal
perturbations in the flow  and for the ensemble response when a flow with a streak
is stochastically excited by temporally and spatially uncorrelated forcing. 

We choose $k_x=3 \alpha$ perturbations. 
This choice of streamwise wavenumber was made because it is the wavenumber that induces the strongest roll circulation
both in DNS and when the flow with the equilibrium  low-speed streak is stochastically excited.
This choice of streamwise wavenumber
for the optimal does not change qualitatively the results that are presented.
The  energy growth  of the first pair of $T=10$  sinuous and varicose  optimals 
is shown in Fig. \ref{fig:Et}  for low-speed streaks (left panel)  and  their mirror high-speed streaks (right panel) with 
with streak amplitude
$\varepsilon=0,\pm0.4,\pm1$. 
This figure shows  that  the optimal perturbation growth  increases as the amplitude of the streak increases and that  the increase is substantial  
when the streak is low-speed and marginal 
when the streak is high-speed. Energy transfer from the mean spanwise shear to the perturbations, $-\int_{\cal D} dy dz~ \overline{uw} U_z$,
is the energy source that accounts for the increased perturbation growth in the presence of the streak, and especially so when a low-speed streak is present because
flows with low-speed streaks have a relatively smaller wall-normal shear and the perturbations are less readily sheared over by the wall-normal shear, which limits their potential growth.

Differences  in the growth of perturbations in flows of
the form  $U_m(y)\pm \varepsilon U_s(y,z)$ is expected, because the flows are 
not  mirror images of each other.  But such pronounced asymmetry in the  growth of optimal perturbations in low-speed 
and their mirror high-speed streaks is surprising and has dynamical implications.
It implies, as we will show, that high-speed streaks are supported weakly by their Reynolds stresses, which provides an  explanation 
for the dominance of low-speed streaks in wall-bounded turbulence.
Fig. \ref{fig:Et}  also shows that in low-speed streak flows the sinuous optimal perturbations achieve substantially 
greater energy growth than their companion varicose perturbations. The structure of the top pair of optimals 
is shown in  Fig. \ref{fig:optk3L}.  As $\varepsilon$ increases the perturbations become increasingly  localized  at the center of the low-speed steak.
Note that the sinuous perturbations  are  concentrated at the wings of the streak,
where there is larger spanwise mean shear. This is consistent with the observation that sinuous perturbations grow more than the 
corresponding varicose perturbation, which  are concentrated at the center of the streak where the shear is small.

\begin{figure}
\centering
\includegraphics[width=0.75\columnwidth]{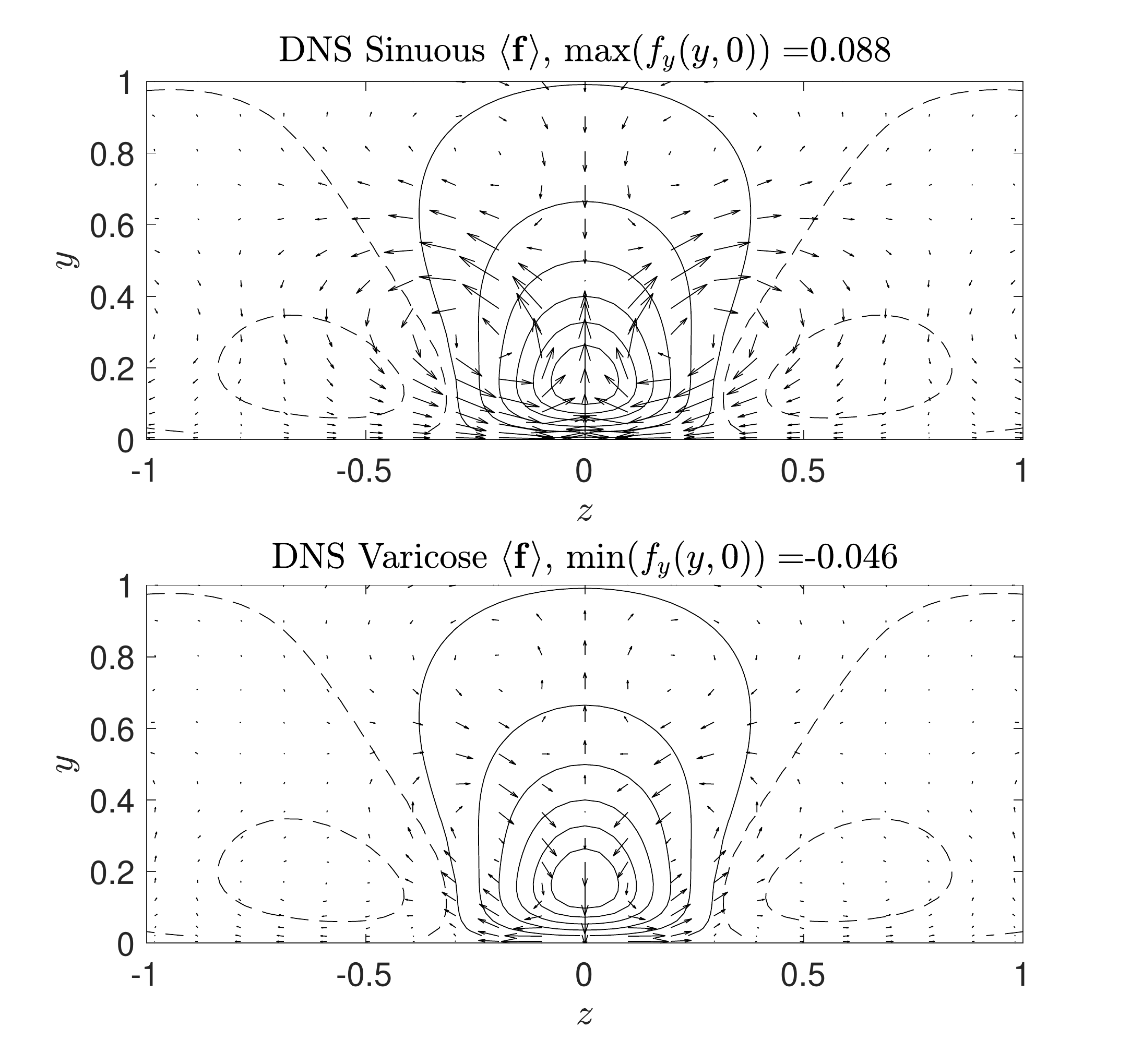}
\caption{Contours of the low-speed streak (solid: negative values, dashed: positive values) and vectors of the  force field, $\langle \mathbf{f} \rangle$, induced by the time mean
Reynolds stresses in the DNS  low-speed streak. Upper panel: the force field induced by the sinuous components of the flow. In agreement with the discussion, they tend to support
the low-speed streak. Lower panel: the force field induced by the varicose components of the flow, which tends to destroy
the low-speed streak.  
The force field of the total perturbation field is shown
in Fig. \ref{fig:stoch_LH_DNS} (upper-panel).}
\label{fig:marios2}
\end{figure}

\begin{figure}
\centering
\includegraphics[width=0.75\columnwidth]{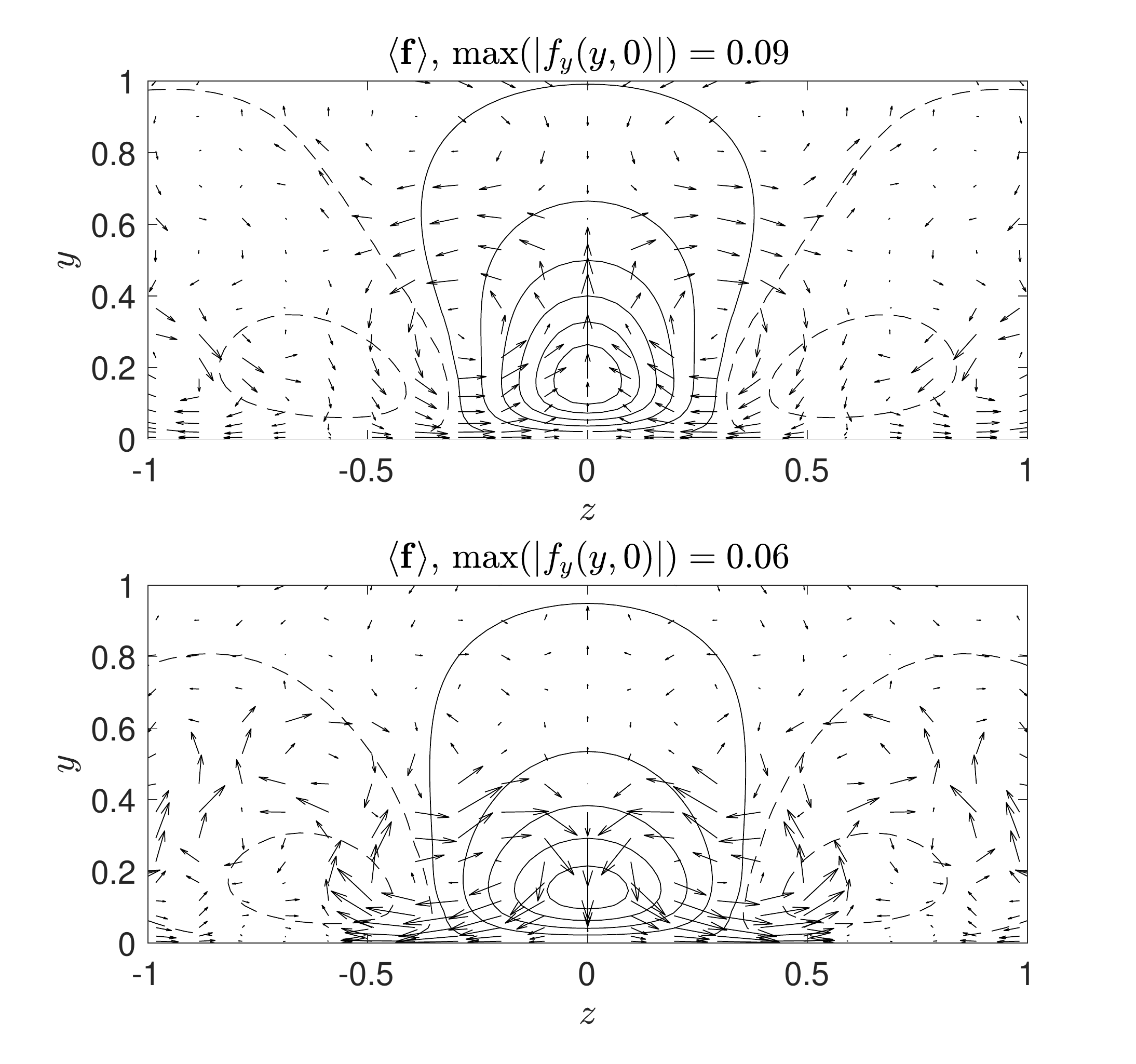}
\caption{Vectors of the force field, $\langle \mathbf{f} \rangle$ induced by the time mean
Reynolds stresses in the DNS  low-speed streak (upper panel) and the high-speed streak (lower panel).
The contours show the respective profile of the average low and high speed streaks in the DNS
Top panel: solid line contours for negative values, dashed for positive values. Bottom panel:
solid line contour for positive values, dashed for positive values. The contour level is 0.03, cf. Fig. \ref{fig:LH}.}
\label{fig:stoch_LH_DNS}
\end{figure}


\begin{figure}
\centering
\includegraphics[width=0.75\columnwidth]{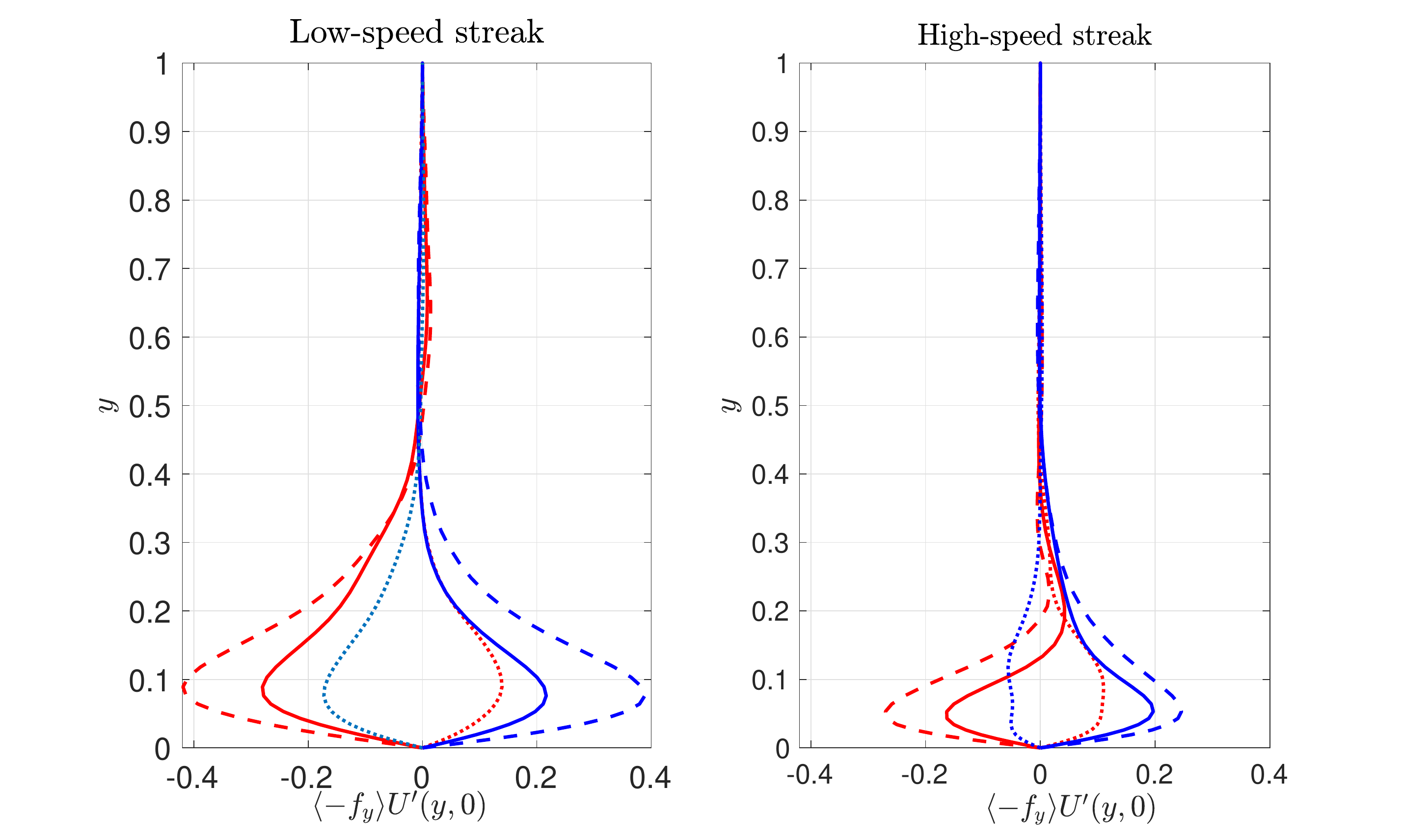}
\caption{The cross-stream distribution  of the net mean acceleration $- f_y  U'(y,0)  $
 from the sinuous (red lines) and varicose (blue lines) $k_x=3 \alpha$ perturbations  in the time-averaged low-speed streak
(left panel) and in the time averaged high-speed streak (right panel).  The streaks are shown in Fig. \ref{fig:LH}. The dashed lines shows the contribution to the induced acceleration
from the Reynolds stress $\overline{v^2-w^2}$  and the dotted lines the contribution from the $\overline{vw}$ Reynolds stress.
The net Reynolds stress  acceleration from  the total perturbation field
tends to support  both the low-speed streak and the high-speed streak. Data from a DNS simulation.} 
\label{fig:marios1}
\end{figure}

%
%
%
%

The time development of the sinuous optimal structure and plots of the resulting rotational Reynolds stress
force field $\fv$ are shown in Fig. \ref{fig:opt_sin_f1}. 
This figure shows that the sinuous optimal induces a coherent roll-circulation 
that tends to reinforce the low-speed streak.  The reverse roll circulation is induced by the companion  varicose optimal, shown  in Fig. \ref{fig:opt_var_f1}.
However,   in the presence of the low-speed streak the streak opposing circulation  of the varicose perturbation is weaker  than the 
streak amplifying circulation of 
the sinuous optimal. Hence, in the presence of even the slightest
streak, the net circulation from  the top pair of optimals tends  to reinforce the low-speed streak, as indicated in Fig. \ref {fig:opt_tot_f1}.
The reverse situation occurs in the high-speed streak. In that case although at $\varepsilon=-1$ the sinuous optimal grows slightly more
than the varicose, the Reynolds stresses from the varicose dominate and  the net circulation tends to reinforce the high-speed streak.
The optimals in this case have very similar structure  (cf. Fig. \ref{fig:optk3H}) and nearly identical evolution and  the net circulation induced
 from the superposition of the opposing and almost identical sinuous and varicose components is weak compared with the circulation induced by the optimals in the low-speed streak flow.

We confirm the robustness of our conclusions about the roll forcing pattern that 
was obtained by analysis of the top pair of optimals 
by considering the general case of roll forcing by the ensemble mean Reynolds stresses  
of  the perturbation field resulting when all degrees of freedom are excited stochastically and equal in energy.
The ensemble mean force field $\fv$ induced when there is  a low-speed streak  present in the flow is shown in Fig. \ref{fig:stoch_LH} (upper-panel),
showing that in low-speed streak flows the sinuous perturbation component dominates the statistics.
In a high-speed streak flow,  Fig. \ref{fig:stoch_LH} (lower-panel),  the induced roll circulation reverses with the varicose modes dominating the statistics.
Note that the induced force is stronger when there is a low-speed streak. The dependence 
of the amplitude of the induced force-field on the amplitude of the streak is plotted in 
Fig. \ref{fig:f_epsilon} from which it is apparent that a  low-speed streak is more vigorously supported by the Reynolds - stresses and this
remains true when the calculation is repeated using the 
high-speed streak with the structure of the equilibrium high-speed streak that that is obtained in DNS (cf. lower panel of Fig. \ref{fig:LH}).
The direction of the induced force, which always tends to increase the pre-existing streak, and the linear dependence of the induced force on streak amplitude,
when the the amplitude of the streak 
is small, indicates that  
the interaction between streaks and  perturbations is consistent with the necessary requirement for producing
an exponential instability. This instability   has been analytically studied
using the S3T form of statistical state dynamics  \citep{Farrell-Ioannou-2012, Farrell-Ioannou-2017-bifur}.

{\color{black}
\section{Comparison with DNS data}
\begin{table}
\begin{center}

\begin{tabular}{@{}*{6}{c}}
\break
 Abbreviation   & $[L_x,L_y,L_z]$ & $[L^+_x,L^+_y,L^+_z]$ &$N_x\times N_z\times N_y$& $R_\tau$& $R$ \\
 NL100   & $[4\pi\;,\;2\;,\;\pi]$ &[1264\;,\;201\;,316] &$128\times 63\times 97$&$100.59$&1650   \\
\end{tabular}
\caption{\label{table:geometry} The simulation of the plane parallel Poiseuille flow was performed in a channel 
of streamwise, wall-normal and spanwise length $[L_x,L_y,L_z]$,  
with periodic boundary conditions in $x$ and $z$ and no slip boundary conditions at the channel walls $y=0$ and $y=2$.
Lengths have been made nondimensional by  $h$ the channel half-width,  velocities by $\langle U \rangle_c$, the center velocity of the time-mean flow, and time by $h/\langle U\rangle_c$. 
The Reynolds number is $R= \langle U \rangle_c h / \nu$, with  $\nu$ the kinematic viscosity.  $[L^+_x,L^+_y,L^+_z]$, indicate
the domain size in wall-units. $N_x$, $N_z$ are the number of Fourier components after dealiasing and $N_y$ is the number of Chebyshev components. $R_\tau= \ut h / \nu$ 
is the Reynolds number of the simulation based on the friction velocity  $\ut= \sqrt{ \nu \left.\df \langle U\rangle/\df y\right|_{\rm w}}$,where $\left.\df \langle U\rangle /\df y\right|_{\rm w}$ is the shear at the wall. }
\end{center}
\end{table}

Observational support for the theoretical arguments that we have presented was obtained in the DNS data  of Poiseuille flow
at $R=1650$. Details of the simulation are given in Table 1. From the DNS 
 we have obtained the average low-speed streak and average high-speed streak, shown in Fig. \ref{fig:LH}, along with their associated perturbation 
 field statistics. 
 The streaks, which were nearly symmetric in the spanwise direction, were  symmetrized and the average Reynolds stresses
 of the sinuous and varicose components of the perturbation field  were obtained. We present here the roll-circulation
 induced by the $k_x=3 \alpha$ component of the perturbation field that produces the largest streamwise-mean torque in DNS.

 The roll-inducing force field $\fv$ produced by the sinuous component of the perturbation field of the low-speed streak is shown in Fig. \ref{fig:marios2}
(upper panel) and that produced by the varicose component in Fig. \ref{fig:marios2} (bottom panel). 
The sinuous component of the DNS tends to amplify
 the low-speed streak, while the varicose tends to induce a weaker reverse circulation, in agreement with the
theoretical discussion in the earlier  sections. In Fig. \ref{fig:stoch_LH_DNS} (top panel) the
roll-inducing component of the force field, $\fv$, produced by the entire  perturbation field is shown, which, 
as discussed above, supports the low-speed streak. 
It was confirmed that the rotational force $\fv$ of the total perturbation field 
 is very close to the sum of the rotational forcing by  the sinuous and varicose components. 
 This indicates that there is no substantial correlation
between the sinuous and varicose components of the flow giving rise to cross-term contributions to  the Reynolds stresses,
so that the total  rotational force $\fv$ and that obtained  from  the analysis performed in this paper, in which the sinuous and 
varicose components independently add, is justified. 
In  Fig. \ref{fig:stoch_LH_DNS} (bottom panel)
is shown the roll-inducing force field $\fv$ of the total perturbation field in the high-speed streak, which 
as expected supports the high-speed streak,
while the support is weaker than that of the $\fv$ in the low-speed streak, as in our previous discussion.

In Fig. \ref{fig:marios1} we plot the contribution of
the  $\overline{v^2-w^2}$ and $\overline{vw}$ Reynolds stresses to the cross-stream acceleration 
of the sinuous and varicose components of the perturbation field in the mean low-speed streak (left panel) and the mean high-speed streak (right panel).
As expected dominant contribution to $\fv$ is confirmed to be due to  $\overline{v^2-w^2}$.
}
%
%
%
%
%
%
%

\section{Discussion}

The study of turbulence dynamics has roots in interpretation of  spatial and temporal spectra 
\citep{Taylor-1935,Taylor-1938,Kolmogorov-1941}.  At the time, data were available from two point space or time correlations which allowed spectra to be determined.  
Unfortunately this required the phase of Fourier components to be left indeterminant and so implicitly the maximum entropy 
assumption of random phase was made.  The result was that turbulence theory neglected the 
role of coherent structures the study of which requires determining the phase of Fourier components.  
The  origin of structures elicited by the observational techniques available was commonly ascribed to modal 
instability based on Rayleigh's theory of modes \citep{Rayleigh-1880,Rayleigh-1896}.  
However, smoke and hydrogen bubbles tended to reveal three dimensional structures that did not correspond with modes of 
maximal growth rate and in fact the canonical laboratory shear flows did not support growing modes.  
The first 3D turbulence to be comprehensively observed was that of the baroclinic turbulence of the midlatitude atmosphere.  
This was accomplished soon after the invention of the vacuum tube which allowed telemetry from weather ballooons to be made over 
the entire troposphere.  Twice daily radiosonde observations covering the entire Northern Hemisphere soon followed.  
The primary coherent structures revealed by this  comprehensive data set was the midlatitude cyclone and the jet stream.  The midlatitude 
cyclone was quickly identified with an unstable mode of the equations of motion linearized about the mean jet flow \citep{Charney-1947,Eady-1949} 
while  the coherent  jet stream structure was determined to result from the cooperative interaction between the 
midlatitude cyclones and the jet, the requisite upgradient momentum flux required for the cyclone 
perturbations to maintain the jet being ascribed to ``negative viscosity'' \citep{Jeffreys-1933,Starr-1953,Starr-1968}.  It is remarkable that the two primary 
mechanisms of coherent structure formation in turbulence, selective  perturbation 
growth  in the linearized equations of motion and cooperative nonlinear interaction 
between essentially stochastic turbulent perturbations and the coherent structure  
had already been recognized within a decade after comprehensive observations of a turbulent flow became available.  
While the fundamental energy source for maintaining the coherent structure by these two dynamical mechanisms
had been identified, the predictions of the analysis of these mechanisms was soon shown to be at variance with observation: 
the growth of cyclones did not correspond with the growth of modes and the  satisfaction of necessary conditions for 
instability did not allow prediction of cyclone formation. In the case of the second mechanism, while  negative viscosity 
provided a suggestive analogy for the upgradient eddy 
momentum fluxes giving rise to the polar jet; as a parameterization in the equations of motion it did not result in 
an analytic theory that predicted jet stream formation in agreement with observations.  Lack of agreement of observed cyclogenesis with
modal theory was subsequently  explained by identification of cyclone formation with non-normal transient growth of stable 
optimally growing perturbations arising in the turbulence \citep{Farrell-1982,Farrell-1984,Farrell-1985,Farrell-1989,Farrell-1989ek}
in which case the dynamics is properly analyzed using generalized stability theory (GST) \citep{Farrell-Ioannou-1996a,Farrell-Ioannou-1996b}.
Understanding the mechanism of jet stream 
formation has roots in the identification of equilibrium structures in stochastic turbulence models  \citep{DelSole-Farrell-1996} 
and obtained full analytic development with the advent of statistical state dynamics methods 
\citep{Farrell-Ioannou-2003-structural,Farrell-Ioannou-2007-structure,Srinivasan-Young-2012,Parker-Krommes-2013,Parker-Krommes-2014-generation,Constantinou-etal-2016,Farrell-etal-2016-PTRSA,Farrell-Ioannou-2019-book}.  
In wall-bounded shear flow turbulence theory the non-normal transient growth mechanism for analyzing 
coherent structure formation was introduced in the case of 2D structures in \cite{Farrell-1988a} and in the case of 3D structures in 
\cite{Butler-Farrell-1992, Butler-Farrell-1993, Farrell-Ioannou-1993a, Farrell-Ioannou-1993b,Reddy-Henningson-1993,Trefethen-etal-1993,Schmid-Henningson-2001}.  
Formation of the R-S by a cooperative instability mechanism was 
advanced  by \cite{Hamilton-etal-1995,Waleffe-1997} and the operative mechanism was identified in \cite{Farrell-Ioannou-2012,Farrell-Ioannou-2017-bifur}.  
More recently methods based on GST have been advanced ascribing the R-S 
to the growth of individual
optimals  \citep{Jimenez-2013lin,Jimenez-2018}
 and alternatively to the excitation of the zero frequency
resolvent  \citep{McKeon-Sharma-2010,McKeon-2017}. 
However, in the case of 3D wall-bounded shear flow it has not been determined whether and in which 
cases the non-normal transient growth mechanism explains
an observed R-S  and in 
which cases the observation is explained by the cooperative instability mechanism, that is whether SSD 
analysis rather than non-normal transient growth analysis using GST is appropriate.

\section{Conclusions}
Our analysis reveals the remarkable universal tendency
of sinuous perturbations  in channel flows to induce 
through their Reynolds stresses roll circulations and also for  varicose perturbations to induce roll circulations of  opposite sign.  This happens even in a stochastically forced channel
with no mean flow. We have ascribed this result to  the presence of solid wall boundaries which
constrain the flow so that the wall normal velocity fluctuations decay to zero faster than the spanwise  
velocity fluctuations. However, when the mean flow  has no spanwise variation the sinuous and varicose perturbations induce
roll
circulations that  cancel each other.
Furthermore, we have shown that even  a streak  at 
perturbative amplitude can organize turbulent perturbation Reynolds stresses so 
as to amplify the perturbing streak which destabilizes the R-S  in 
turbulent shear flows.  The reason for this universal streak amplification property 
is revealed by the partitioning of the turbulent perturbation field into sinuous and 
varicose components.  With this partition the breaking of the tendency for 
cancellation of the opposing sinuous and varicose structure Reynolds stresses 
in a spanwise homogeneous perturbation field is found to be broken by the imposition 
of the streak resulting  in the sinuous component dominating in the case of a 
low speed streak and the varicose component  dominating in the case of a high 
speed streak. This result when coupled with the fact that the sinuous perturbations 
are favored  in growth by the low speed streak while the varicose are favored 
by the high speed streak  provides an analytic explanation for the universal 
streak amplification property in shear flow.   We find also that 
the Reynolds stress forcing of the roll circulations of low-speed streaks is stronger than that of high-speed streaks, indicating that 
a contributing factor for the observed relative weakness of high-speed streaks in wall-turbulence is the weaker forcing
by the Reynolds stresses that the high speed streak organizes.

As the origin and maintenance of the R-S is central to 
the theory of turbulence in shear flow, in this work we have performed an in depth  analysis of the physical mechanism 
by which the R-S arises in Poisseullle flow at $R=1650$.
We find that the R-S arises by organization of 
Reynolds stresses producing roll inducing torques resulting in an SSP  in agreement with predictions of
the cooperative SSD mechanism.  
In addition we have shown that turbulent Reynolds stresses 
as predicted and required by this SSD cooperative perturbation-roll-streak mechanism are observed to be
in agreement with DNS data of turbulence of the same flow which supports the conclusion that this mechanism we have 
identified  continues  at finite amplitude to 
support the maintenance of the turbulent state.

\bibliographystyle{jfm}

\end{document}